\documentclass[preprint,amssymb,amsmath,amsfonts,nobibnotes,footinbib]{revtex4}

\usepackage{blkarray}

\usepackage[utf8]{inputenc}
\usepackage[T1]{fontenc}
\usepackage[colorlinks=true,linkcolor=red,pdfencoding=auto,psdextra]{hyperref}

\usepackage{graphicx}
\usepackage[font=small]{caption}

\usepackage{bm}
\usepackage{dsfont}
\usepackage{amsthm}

\begin{document}

\newcommand{\textred}{\textcolor{red}}

\newcommand{\diag}{\textup{diag}} 
\newcommand{\Mean}{\mathbb{E}} 
\newcommand{\prob}[1]{\mathbb{P}\left\{#1\right\}} 
\newcommand{\Eye}{\mathds{1}} 
\newcommand{\wt}[1]{\widetilde{#1}}
\newcommand{\allone}{\bm{1}}
\newcommand{\bpi}{\bm{\pi}}
\newcommand{\bPi}{\bm{\Pi}}
\newcommand{\bq}{{\bf q}}
\newcommand{\bQ}{{\bf Q}}
\newcommand{\bA}{{\bf A}}
\newcommand{\bC}{{\bf C}}
\newcommand{\veps}{\varepsilon}

\title{Exact and Approximate Mean First Passage Times on Trees and other Necklace Structures: a Local Equilibrium Approach}

\author{Yanik-Pascal F\"{o}rster}
\author{Luca Gamberi}
\author{Evan Tzanis}
\author{Pierpaolo Vivo}
\author{Alessia Annibale}

\affiliation{Quantitative and Digital Law Lab, Department of Mathematics, King's College London,
	Strand, WC2R 2LS, London (United Kingdom)}

\homepage[]{Quantlaw.co.uk}

\date{\today}

\begin{abstract}

In this work we propose a novel method to calculate 
mean first-passage times
(MFPTs) for random walks on graphs, based on a 
dimensionality reduction technique for Markov State Models, known as local-equilibrium (LE). 
We show that for a broad class of graphs, 
which includes trees, 
LE coarse-graining preserves the MFPTs between certain nodes, upon making a suitable choice of the coarse-grained states (or clusters). 
We prove that this relation is exact for graphs that can be coarse-grained into a one-dimensional lattice where each cluster connects to the lattice only through a single node of the original graph. 
A side result of the proof generalises the well-known essential edge lemma (EEL), which is valid for reversible random walks, to irreversible walkers. Such {a} generalised EEL leads to explicit formulae for the MFPTs between certain nodes in this class of graphs.
For graphs that do not fall in this class, the generalised EEL provides useful approximations if 
the graph allows a one-dimensional coarse-grained representation and the clusters are sparsely interconnected.

We first demonstrate our method for the simple random walk on the $c$-ary tree, then we consider other graph structures and more general random walks, 
including irreversible random walks.

\end{abstract}

\maketitle

\section{Introduction\label{sec:intro}}

Random walks on networks are intuitive and highly general 
stochastic processes
that enjoy attention in many different applications. Examples include models for foraging and predator-prey behaviour \cite{Humphries2010,Viswanathan1999} among numerous other biological applications \cite{Codling2008}, centrality measures such as the famous PageRank \cite{Brin1998,Page1998}, and search strategies \cite{Benichou2014} as in hide-and-seek games \cite{Pandey2018}. As a consequence, random walks on networks, especially the \emph{simple} random walk, where the hopping probabilities from any network node to all adjacent nodes are uniform, are well studied. 

First-passage times (FPTs), i.e. times at which certain events occur for the first time, are important observables of many stochastic and, in particular, Markov processes, with random walks being no exception. {Consider for instance the Gambler's ruin (the first time a Gambler's budget hits zero) or break-even points in trading (when the selling price of a stock exceeds the price paid for the first time) \cite{Feller1968}, and extinction events in birth-death-processes \cite{Doering2005} (see \cite{Redner1951,Aldous1989,Aldous1999} for further examples).}
The set of \emph{mean} first passage times (MFPTs) between the states of a Markov process 
encapsulates fundamental properties of the system's kinetics 
via their relation to the spectrum and eigenvectors of the transition matrix \cite{Redner1951,Lovasz1994,Kells2020} and the relaxation times of the random walker which are all of particular computational importance \cite{Noh2004}. MFPTs have also been shown to provide important information about 
correlations and heterogeneity in complex systems \cite{Bassolas2020},
and optimal 
coarse-graining in Markov State Models \cite{Kells2020}.


MFPTs of random walks on networks
encode \emph{global} properties of the random walkers and 
the network they explore, hence their explicit and exact calculation can in general be very hard for networks larger than a few nodes. There are many ways to express the full matrix of MFPTs theoretically; one of the classical and most general methods is to employ the so-called fundamental matrix, as proposed in \cite{Kemeny1960}. The fundamental matrix is also connected to equilibrium properties and commute times of the walker and has been revisited and reformulated over time, for instance in \cite{Meyer1975}. 
Reversible random walkers, where hopping probabilities are in detailed balance with the equilibrium node-occupancy probabilities, are often more accessible. Here, some popular approaches include the network analogue of resistance theory (see \cite{Bapat2011} for application to trees and \cite{Redner1951} for a general introduction) and the essential edge lemma, which applies when the graph consists of 
subgraphs that are connected by a single edge (see e.g. \cite{Aldous1999}).

These exact methods rarely lead to explicit results even in simple cases, when one would hope to express MFPTs e.g. in terms of the graph parameters of a model. {Only for very specific problems, e.g. in the presence of a high degree of symmetry and hierarchy, the problem can be solved explicitly by successive ``decimation'' procedures \cite{VanDenBroeck1989, Agliari2008, Balakrishnan2019}.}
This has led to the development of various approximation schemes -- for instance, mean-field approaches based on node degree \cite{Baronchelli2008}, or on the distance from a target \cite{Baronchelli2006}. For sufficiently dense networks with random weights, the information contained in the neighbourhood of the target node is sufficient to formulate an accurate rank-$1$ approximation for the MFPT from any other node \cite{Bartolucci2021}. 
For sparser networks, the approach presented in \cite{Martin2010} exploits locally tree-like structures 
to derive asymptotic expressions for a large number of nodes. 
Moreover, tail-estimates for first-passages of rare events can be constructed for many different applications \cite{Aldous1989}.
The approximations made in these works tend to be valid either in the limit of large graphs, $n\to\infty$, or are restrictive about the type of random walker to which they can be applied. For example, many are developed for {\emph{simple} (purely diffusive) random walks.} 
For more general dynamics, explicit results are hardly available.

In this paper, we show that kinetic coarse-graining techniques, introduced to reduce the dimensionality of Markov State Models, can be used to derive \emph{explicit} formulae for MFPTs in terms of the graph parameters. These are exact for a broad class of graphs, which includes tree-graphs.

The method is based on three key ideas: (i) upon kinetically coarse-graining random walks on graphs, the calculation of MFPTs simplifies due to the reduced dimensionality of the coarse-grained system, (ii) 
for certain graph structures,
it is possible to adopt coarse-grained representations 
that drastically simplify the calculations, leading to explicit formulae for the MFPTs in the coarse-grained space, 
(iii) for these graph structures, under some conditions, the MFPTs of the coarse-grained system match exactly certain MFPTs in the original system{.}
When these conditions are violated, the MFPTs of the coarse-grained random walk may still provide reliable approximations of certain MFPTs in the original system.

In particular, we prove that in 
graphs with special, i.e. ``necklace'', topologies
certain MFPTs in the original dynamics can be calculated exactly using a coarse-graining technique  
known as \emph{local-equilibrium} (LE). Our proof  
is valid for general random walks, including irreversible random walks, 
as long as they converge to a steady-state distribution. In addition to this proof, which leads to  eq.~\eqref{eq:mfpt_LE_conservation_all}, our analysis provides two main results. The first one, eq.~\eqref{eq:essential_edge_lemma_general}, is a generalisation of the popular essential edge lemma (EEL) \cite{Aldous1999}, which is only valid for reversible random walks{; it} is retrieved here  
as a special case of a more general equation, which does not require dynamical reversibility. This result leads to explicit formulae for MFPTs in simple random walks on graphs with necklace
topologies. 

The second main result, eq.~\eqref{eq:mfpt_LE_derv_n_1}, 
provides a method 
that allows getting reliable approximations of MFPTs, using LE coarse-graining,
for more general random walks and graph topologies, as long as they can be coarse-grained into a one-dimensional lattice. The applicability of this (approximate) method is significantly broader than the generalised EEL, eq.~\eqref{eq:essential_edge_lemma_general}.

In sec.~\ref{sec:definitions} we review the notions of MFPTs and LE coarse-graining. 
In sec.~\ref{sec:main_derivation} we show that LE coarse-graining preserves certain MFPTs for general random walks on a broad class of graphs with ``necklace'' structure and we provide a generalization of the essential edge lemma, valid for reversible random walks, to irreversible random walks. In sec.~\ref{sec:mfpt_tree} we demonstrate how the LE coarse-graining method can be used to derive explicit MFPTs formulae for the simple random walk on $c$-ary trees. These expressions are consistent with those resulting from the essential edge lemma.
In sec.~\ref{sec:applications_rev} 
we apply the generalised essential edge lemma, derived in {sec.}~\ref{sec:main_derivation}, to simple random walks on 
non-tree graphs with necklace structures. In sec.~\ref{sec:applications_non_rev}, we apply the method to irreversible random walkers, 
where the popular EEL does not apply.
Sec.~\ref{sec:LE_approx_for_mfpts}  
considers
two scenarios in which exact results are more difficult 
to obtain, whereas coarse-graining leads to explicit, approximate formulae for MFPTs. We summarise results in sec.~\ref{sec:conclusion}. 
Some of the technical definitions of our derivations are elaborated in app.~\ref{app:spanning_trees}. {In app.~\ref{app:higher_moments} we briefly explore the possibility to extend our results to higher moments and full distributions of FPTs.}

\section{Definitions}
\label{sec:definitions}
We consider random walkers on $n$-node graphs, with vertices labelled as $i,j{,}\ldots$ and transition probability matrix $\bm{q}$, such that the element $q_{ij}$ denotes the probability to go from node $i$ to node $j$ in one time step. If there is no edge from node $i$ to node $j$, then $q_{ij}=0$, hence the matrix $\bm{q}$ defines a directed graph where every directed edge $(i,j)$ is weighted by the hopping probability $q_{ij}$ of the walker. From normalisation of probabilities, one has $\sum_{j=1}^n q_{ij}=1~\forall ~i$.

We will focus on irreducible random walks, for which the set of MFPTs $m_{ij}$ from any site $i$ to any site $j$ is determined by the recurrence equations
\begin{equation}\label{eq:mfpt_basic_recurrence}
    m_{ij} = q_{ij}+\sum_{k\neq j} q_{ik}(m_{kj}+1)\ ,
\end{equation}
where the first term 
accounts for the walker hopping from $i$ to $j$ directly (which occurs with probability $q_{ij}$), while the second term accounts for the walker hopping to any other node $k$ first and 
starting a first-passage process from there (at the next time step). Using normalisation of ${\bf q}$,  eq.~\eqref{eq:mfpt_basic_recurrence} can be rearranged into an expression for the vector $\bm{m}_j$ of MFPTs to $j$ starting from all other vertices, given by \cite{Masuda2017}
\begin{equation}\label{eq:mfpt_grounded_laplacian}
    \bm{m}_j = \left( \Eye_{n-1} - \widehat{\bm{q}}_{j} \right)^{-1} \allone_{n-1}\ ,
\end{equation}
where $\Eye_n$ and $\allone_n$ are the identity matrix 
and the all-$1$ vector of size $n$, respectively, and  
$\widehat{\bm{q}}_{j}$ is the transition matrix of the walker from which the $j$-th row and column have been removed.  
Here and below, 
we think of vectors as columns, referring to row vectors as the transpose $(-)^{T}$ of a column. 
We define by $\bpi^T$ the row vector of steady-state probabilities, such that $\bpi^T \bq=\bpi^T$. Due to our assumption that the random walker is irreducible, the steady-state is unique.
We recall that when the dynamics {are} reversible, i.e. the detailed balance condition 
\begin{align}\label{eq:DB}
    \pi_{i}q_{ij} = \pi_{j}q_{ji}
\end{align}
is satisfied for any pair of vertices $i,j$, the steady-state is the equilibrium state. For reversible random walks, a number of exact methods to obtain MFPTs do exist. However, in this work, we will not assume that eq.~\eqref{eq:DB} is satisfied. We will simply assume that the system converges to a unique steady-state.

Given a random walk ${\bf q}$ on $n$ nodes, one can define a \emph{coarse-grained} random walk $\bQ$ on $N$ nodes, where $N<n$, by grouping together the nodes $i,j, \ldots$ of the original network 
into $N$ subgraphs, or \emph{clusters}, labelled by upper case indices $I, J, \ldots$. 
This operation can be encoded 
into an $n\times N$ matrix $\bC$, whose elements $C_{iI} \in \{1,0\}$ denote whether $(1)$ or not $(0)$ node $i$ belongs to subgraph $I$, for all $i=1,\ldots,n$ and $I=1,\ldots,N$.

There has been much recent research into how to optimally define the transition matrix ${\bf Q}$ of hopping probabilities between clusters,  
for a given choice of the clustering $\bC$ \cite{Hummer2015}. 
For $\bQ$ to retain the 
equilibrium properties of the original dynamics, its left eigenvector associated to the unit eigenvalue must satisfy $\bPi^T=\bpi^T \bC$, i.e. 
the steady-state occupancy probability of a cluster 
must equate the sum of the steady-state occupancy probabilities of the nodes in that cluster. 
This, however, does not determine $\bQ$ uniquely and 
further conditions must be imposed. A popular prescription, known in the literature as the  \emph{local-equilibrium} (LE) clustering \cite{Hummer2015}, requires that the probability flux from cluster $I$ to cluster $J$ be equal to the sum of the probability fluxes from any node $i$ in cluster $I$ to any node $j$ in cluster $J$
\begin{equation}\label{eq:LE_def}
    Q_{IJ}=\frac{1}{\Pi_I}\sum_{ij} C_{iI}\pi_i q_{ij}C_{jJ}\ .
\end{equation}
Due to the reduced dimensionality of $\bQ$, when compared to $\bq$, certain observables may be easier to 
calculate in the coarse-grained graph. 

In the next section we focus on a broad class of graphs where subgraphs can be arranged in a line, 
such that each subgraph connects to the line only through one vertex. We prove that the MFPTs between these vertices in the original dynamics are equal to the 
MFPTs between the corresponding clusters in the LE coarse-grained dynamics, for which we are able to derive explicit formulae.
The result is general, in particular it does not require reversible dynamics.
Hence, our method provides quick access to an explicit MFPT formula whenever information on the steady-state cluster occupancy probability $\bPi$ is available.

\section{Conservation of MFPTs under coarse-graining\label{sec:main_derivation}}

\begin{figure}[h]
    \centering
    \includegraphics{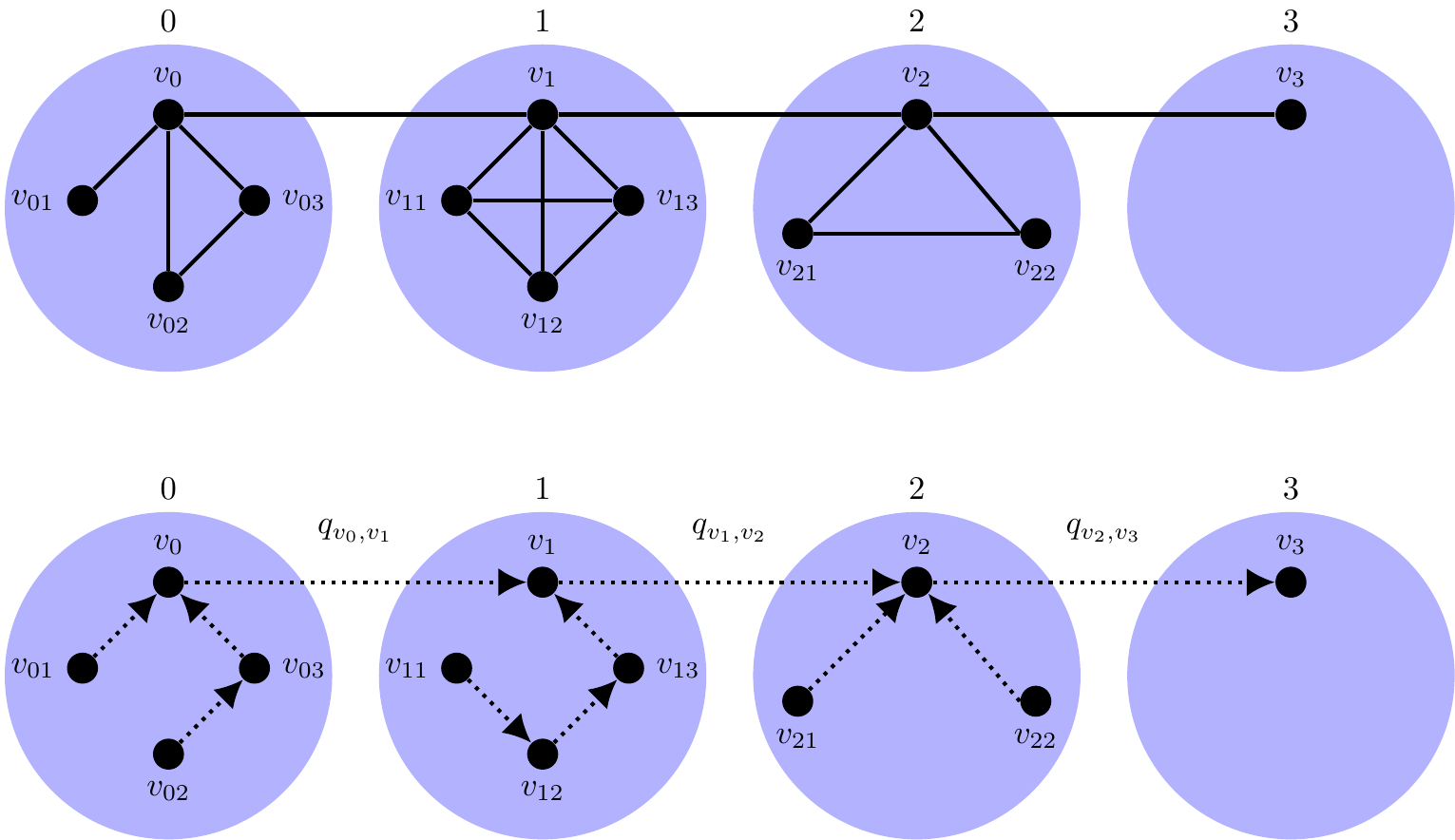}
    \caption{Top: Example of a necklace with $H=3$. Bottom: Dotted arrows form a spanning tree with root $v_3$. Since the subgraph-interconnections consist of single edges, every such spanning tree must contain the edges $(v_0,v_1)$, $(v_1,v_2)$ and $(v_2,v_3)$, with weight {$q_{v_0, v_1}$, $q_{v_1, v_2}$ and $q_{v_2, v_3}$}, respectively.}
    \label{fig:spanning_trees_example}
\end{figure}

In this section, we prove that coarse-graining according to LE preserves certain MFPTs of the random walker exactly, if the graph has a special ``necklace'' structure, i.e. if it can be regarded as a {one-dimensional} ``chain of graphs''.
More precisely, we consider  
graphs consisting of $H+1$ disjoint, connected subgraphs $0,\dots,H$ 
hanging from the line $v_0,\dots,v_H$ (the \emph{backbone}) of distinguished vertices $v_I \in I$, with $I=0,\ldots,H$; see the top of fig.~\ref{fig:spanning_trees_example} for an illustration. The subgraphs can have arbitrary structure, as long as their only interconnections are the links in the backbone. 
We are interested in MFPTs to the target node $v_H$, from another node in the backbone, initially set to $v_0$. 
Without loss of generality, we define the target subgraph as containing only the target node, $H=\{v_H\}$, as we are only concerned with MFPTs to $v_H$ from outside $H$.
We denote vertices \emph{within} the subgraph $I$ other than $v_I$ by $v_{Ii}$ with $i=1,\ldots ,|I|-1$ {, where $|I|$ is the number of vertices in $I$}.
To keep the notation uniform, 
we denote $v_I$ as $v_{I0}$ if necessary.

In accordance with the notation introduced in the  section above, we will use small letters -- e.g. $\bm{q}$, $m_{ij}$, $\bm{\pi}$ -- to refer to properties of the original random walker, and capital letters -- $\bm{Q}$, $M_{IJ}$, $\bm{\Pi}$ -- for corresponding properties of the coarse-grained walker.

In the following we derive a general formula 
for the MFPT $m_{v_0 v_H}$ from $v_0$ to $v_H$ that outmanoeuvres the 
inversion formula \eqref{eq:mfpt_grounded_laplacian}, 
and we demonstrate that this matches exactly 
the MFPT $M_{0 H}$ of a walker in the coarse-grained graph, where each subgraph is regarded as a cluster, 
and the hopping probabilities between clusters are 
defined according to the LE prescription.
Results generalise immediately to arbitrary pairs of nodes in the backbone. 

To the best of our knowledge, our result is the first to lead to explicit and exact formulae for MFPTs in graphs with necklace structure. This is a broad class of graphs, which includes tree-graphs, as we will show in the next section.
Similar graph structures were considered in \cite{Matan1989}, where expected escape-times from clusters connected to a one-dimensional lattice each by a single edge, were calculated. 
This approach, however, did not rely on coarse-graining techniques, which 
broaden the practical use of our formula.

\begin{figure}[h]
    \centering
    \includegraphics{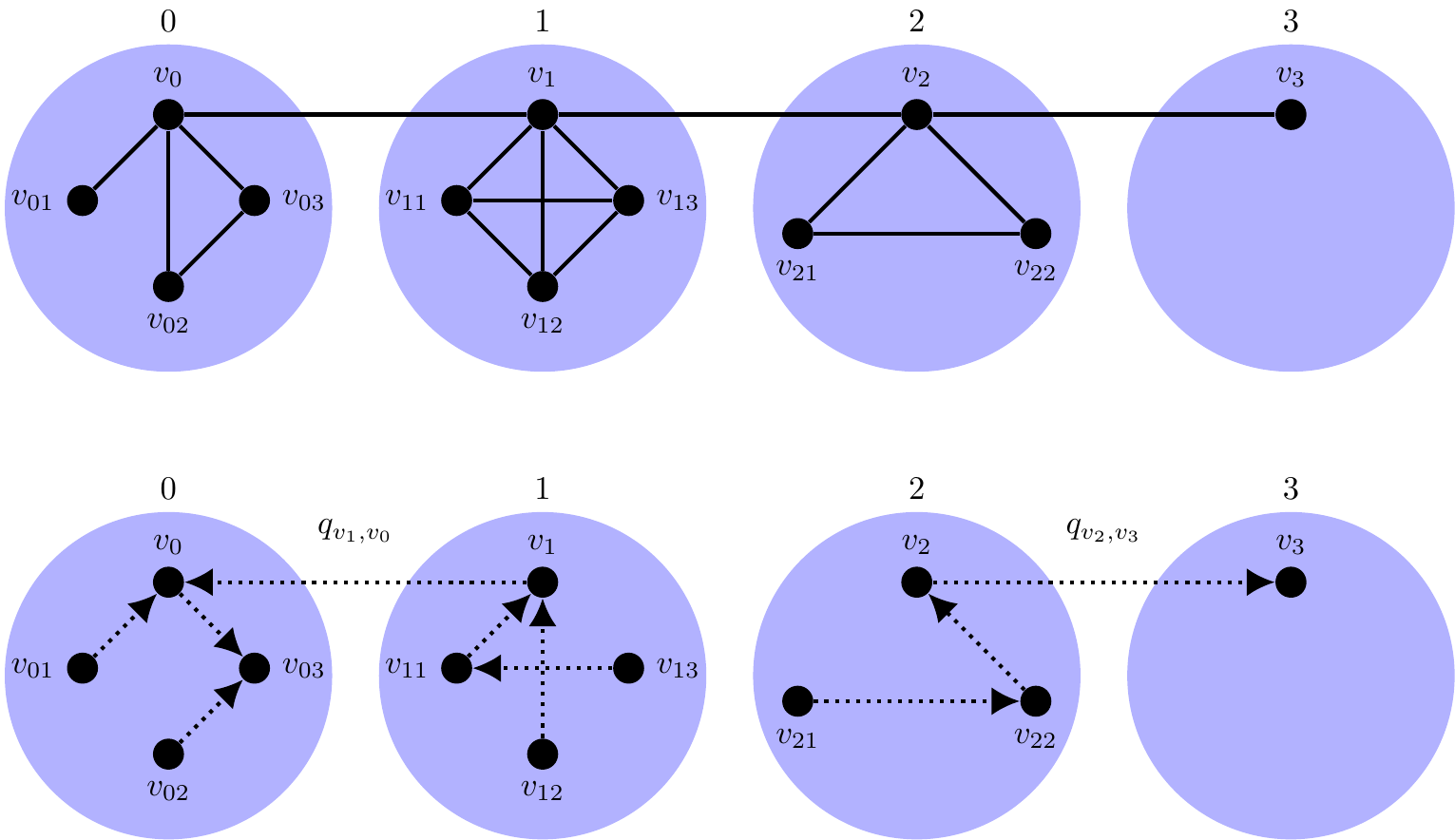}
    \caption{Top: Example of a necklace with $H=3$, as in fig.~\ref{fig:spanning_trees_example}. Bottom: Dotted arrows show a two-tree forest with one component rooted in $v_3$, the other containing $v_0$. In this case, the $v_0$-component has root $v_{03}$. For a $v_0$-component with any other root in subgraph $0$, all edges outside $0$ stay the same, as any path coming in to $0$ must go through $v_0$.}
    \label{fig:two_forest_example}
\end{figure}

Our proof relies on a combinatorial approach to the calculation of MFPTs, which consists in finding all the 
spanning trees and two-tree forests 
in the graph where the random walk takes place (see app.~\ref{app:spanning_trees} for the definition of spanning trees and forests).
Upon defining the weight of a tree $\mathfrak{t}$ as the product of all its edge weights 
\begin{equation}
    w(\mathfrak{t}) = \prod_{(ij)\in \mathfrak{t}}q_{ij}\ ,
\end{equation}
where the product runs over the edges of the tree, the MFPT from node $i$ to $j$ is found within this approach as  \cite{Chebotarev2007,Pitman2018}
\begin{equation}\label{eq:mfpt_tree_formula}
    m_{ij} = \frac{s_{ij}}{s_j}\ .
\end{equation}
Here $s_j$ is the sum of the weights of all spanning trees rooted in $j$, which we denote by $\mathfrak{t}\to j$ (as, by definition of root, all edges ``point toward'' the root){,}
\begin{equation}\label{eq:single_root_weight}
    s_j = \sum_{\mathfrak{t}\to j} w(\mathfrak{t})\ ,
\end{equation}
and $s_{ij}$ is the sum of the weights of all two-tree forests $(\mathfrak{t},\mathfrak{s})$, such that $\mathfrak{t}$ has root $j$ and $\mathfrak{s}$ contains $i$ (but can have any root; cf. fig.~\ref{fig:two_forest_example} and further examples in app.~\ref{app:spanning_trees})
\begin{equation}\label{eq:two_root_weight}
    s_{ij} = \sum_{\mathfrak{t}\to j; i\in\mathfrak{s}} w(\mathfrak{t})w(\mathfrak{s})\ .
\end{equation}
Conveniently, one can express the stationary probabilities of an irreducible random walker in terms of the same quantities \cite{Aldous1999}
\begin{equation}\label{eq:MC_tree_formula}
    \pi_{j}=\frac{s_j}{\sum_{k=1}^n s_k}\ .
\end{equation}

We apply eq.~\eqref{eq:mfpt_tree_formula} to the sites $v_0$ and $v_H$. For graphs with necklace structure, as shown in fig. \ref{fig:spanning_trees_example}, any spanning tree with root $v_H$ must contain all edges of the backbone pointing in the direction of $v_H$, as these are \emph{essential edges}, i.e. the graph becomes disconnected when any one of them is removed.
This means that the path weight $q_{v_0 v_1}\cdot\dots \cdot q_{v_{H-1} v_H}$ is a common factor in the sum in eq.~\eqref{eq:single_root_weight}. We will use the shorthand 
\begin{equation}\label{eq:path_weight}
    w_{IJ} = 
    \begin{cases}
    \prod_{K=I}^{J-1} q_{v_K v_{K+1}} \quad &\colon \textup{if } I<J\, \\ 
    \prod_{K=J}^{I-1} q_{v_{K+1} v_K} \quad &\colon \textup{if } I>J\, \\ 
    1 \quad &\colon \textup{else}\,
    \end{cases}
\end{equation}
for products of the hopping probabilities $q_{v_I v_J}$,
and $W_{I J}$ for the corresponding products of the hopping probabilities $Q_{I J}$ between clusters.

Furthermore, every subgraph $I$ is connected to the backbone only via $v_I$, forcing the sub-spanning trees within each $I$ to be rooted in $v_I$. Denoting the sum of weights of all sub-spanning trees of $I$ by $w(I\to v_I)$, 
and noting that the backbone and the attached subgraphs account for all of the vertices in the graph, we can write $s_{v_H}$ in the 
factorised form
\begin{equation}\label{eq:spanning_tree_factorised}
    s_{v_H} = w_{0H} \prod_{I=0}^H w(I\to v_I)\ .
\end{equation}

Similarly, we can decompose the two-tree forest weight $s_{v_0 v_H}$ defined in eq.~\eqref{eq:two_root_weight}. 
Each relevant two-tree forest consists of a tree rooted in $v_H$, 
and a tree that contains $v_0$ but can be rooted in any of its vertices, including vertices that are not on the backbone. 
Firstly, we observe that whenever $v_H$ and $v_0$ lie in different trees of a spanning forest, one of the (undirected) edges of the backbone, say $\left(v_{I-1}, v_I\right)$ with $I>1$, must have been omitted in the forest, for example in fig.~\ref{fig:two_forest_example}, it is $(v_1,v_2)$. 
All other edges of the backbone must necessarily be included, each in one direction.
Consequently, the $v_H$-component contributes a weight $w_{IH}\prod_{J=I}^H w(J\to v_J)$ for the forest in which subgraphs $I$, $I+1$, \dots, $H$ are included in the $v_H$-component. The other component contributes three factors: 
\begin{enumerate}
    \item for any fixed $K=0,\dots ,I-1$ there is a weight $w_{0K}w_{I-1,K}$ for the backbone edges pointing towards subgraph $K$,
    \item for any vertex $v_{Km} \in K$ with $K$ fixed as above, we have a weight $w(K\to v_{Km})$ for the spanning trees of $K$ pointing towards the vertex,
    \item the remaining subgraphs $J=0,\dots,\ K-1,\ K+1,\ \dots, I-1$ give rise to $w(J\to v_J)$.
\end{enumerate}
For instance, the $v_0$-component in fig.~\ref{fig:two_forest_example} is rooted in the node $v_{03}$ of $0$ and has weight $w(0\to v_{03})w(1\to v_1)q_{v_1 v_0}$; the other component has weight $w(2\to v_2)w(3\to v_3)q_{v_2 v_3}$. 
Summing the product of these weights over $I$, $J$, $K$ and $m$, 
one obtains
\begin{align}\label{eq:spanning_forest_factorised}
\nonumber    s_{v_0 v_H}
    &=\sum_{I=1}^{H}\left[ w_{I H} \prod_{J=I}^H w(J\to v_J) \right. \\ \nonumber
    &\left. \times \sum_{K= 0}^{I-1}\left( \sum_{m=0}^{|K|-1} w(K\to v_{Km}) w_{0 K}\, w_{I-1,K}
     \prod_{J= 0,J\neq K}^{I-1} w(J\to v_J)  \right) \right] \\ 
    &= \sum_{I=1}^{H}\sum_{K=0}^{I-1} \left[ w_{I H}\, w_{0 K}\, w_{I-1,K} \frac{\prod_{J=0}^H w(J\to v_J)}{w(K\to v_K)}  \sum_{m=0}^{|K|-1}w(K\to v_{Km})  \right]\ .
\end{align}
To apply eq.~\eqref{eq:mfpt_tree_formula}, we divide this expression by the one in eq.~\eqref{eq:spanning_tree_factorised}. Due to the factorisation $w_{0H}=w_{0K}w_{KI}w_{IH}$ implied by eq.~\eqref{eq:path_weight}, the factors $w_{0K}$ and $w_{IH}$ in the numerator are cancelled by $w_{0H}$; we therefore arrive at the simplified formula
\begin{align}\label{eq:essential_edge_lemma_general}
    \nonumber m_{v_0 v_H} &= \sum_{I=1}^{H}\sum_{K=0}^{I-1} \frac{ w_{I H}\, w_{0 K}\, w_{I-1,K} }{  w_{0 H} \prod_{J=0}^H w(J\to v_J)} \frac{ \prod_{J=0}^H w(J\to v_J) }{ w(K\to v_K) } \sum_{m=0}^{|K|-1} w(K\to v_{Km}) \\
    &=\sum_{I=1}^{H}\sum_{K=0}^{I-1} \frac{ w_{I-1,K} }{w_{K I}} \frac{\Pi_K}{\pi_{v_K}}\ ,
\end{align}
where we have written $\bm{\pi}$ in terms of tree weights via eq.~\eqref{eq:MC_tree_formula} and used the definition $\Pi_K=\sum_{ m=0 }^{|K|-1}\pi_{v_{Km}}$ for the equilibrium cluster occupancy probability of the random walker. Eq.~\eqref{eq:essential_edge_lemma_general} is a useful result in its own right, which we demonstrate in the following sections. Due to its connection to the EEL shown at the end of this section, we refer to eq.~\eqref{eq:essential_edge_lemma_general} as \emph{generalised essential edge lemma} (GEEL).

We now {make} the same simplification for the coarse-grained walker. Retaining our convention of using capital letters for reference to coarse-grained dynamics, we are interested in 
\begin{equation}
    M_{0 H} = \frac{S_{0 H}}{S_{H}}\ .
\end{equation}
Upon choosing the subgraphs $I=0,\ldots,H$ as the 
clusters of the coarse-grained dynamics, these are collapsed into single vertices, and all spanning trees become lines (see as illustrations  figs.~\ref{fig:spanning_trees_example} and \ref{fig:two_forest_example} where all nodes within the same shaded area are identified).
Consequently, $S_{H}$ is the weight of the directed path from $0$ to $H$, 
\begin{equation}\label{eq:LE_spanning_tree_factorised}
    S_{H}= \prod_{I=0}^{H-1} Q_{I,I+1}=W_{0 H}\ .
\end{equation}
On the other hand, all two-tree forests contributing to $S_{0 H}$ can again be found by omitting one edge of the path $0, \dots, H$, directing the $H$-component towards $H$ and having the $0$-component point anywhere. As above, these two requirements can be condensed into
\begin{equation}\label{eq:LE_spanning_forest_factorised}
    S_{0 H}=\sum_{I=1}^{H}\sum_{K=0}^{I-1}  W_{0 K}\, W_{I-1,K}\, W_{I H}\ . 
\end{equation}
In the quotient of eqs.~\eqref{eq:LE_spanning_tree_factorised} and \eqref{eq:LE_spanning_forest_factorised}, we can again use the factorisation $W_{0H}=W_{0K}W_{KI}W_{IH}$ in the denominator as above, such that the term $W_{0K}W_{IH}$ is cancelled. Thus eq.~\eqref{eq:mfpt_tree_formula} becomes
\begin{equation}\label{eq:mfpt_LE_derv_n_1}
    M_{0 H}=\sum_{I=1}^{H}\sum_{K=0}^{I-1} \frac{ W_{I H}\, W_{0 K}\,  W_{I-1,K}}{W_{0 H}} 
    = \sum_{I=1}^{H}\sum_{K=0}^{I-1} \frac{W_{I-1,K}}{W_{K I}}\ .
\end{equation}
For the comparison with eq.~\eqref{eq:essential_edge_lemma_general}, we express the LE path weights $W$ in terms of the unclustered weights $w$. To this end, we apply the LE definition of coarse-graining, eq.~\eqref{eq:LE_def},
and the chain structure of the coarse-grained graph, 
which implies
\begin{equation}
    Q_{I J}=\left(\delta_{I,J+1} + \delta_{I,J-1}\right)\frac{\pi_{v_I}}{\Pi_I}q_{v_I v_J}\ .
\end{equation}
Substituting the transition probabilities of the coarse-grained dynamics in the definition of $W_{I J}$ given in eq.~\eqref{eq:path_weight}, we can rewrite the fraction in eq.~\eqref{eq:mfpt_LE_derv_n_1} as follows
\begin{align}\label{eq:mfpt_LE_derv_n_2}
\nonumber    \frac{W_{I-1,K}}{W_{K I}} &=\left[\prod_{J=K+1}^{I-2} \frac{Q_{J,J-1}}{Q_{J,J+1}}\right]  \frac{Q_{I-1,I-2}}{Q_{K,K+1}Q_{I-1,I}} = \left[\prod_{J=K+1}^{I-2} \frac{q_{v_{J,J-1}}}{q_{v_{J,J+1}}}\frac{q_{v_{I-1,I-2}}}{q_{v_{K,K+1}}}  \right]  \frac{\Pi_K}{q_{v_{I-1,I}} \pi_{v_K} } \\ 
    &=  \frac{ w_{I-1,K} }{w_{K I}} \frac{\Pi_K}{\pi_{v_K}}\ .
\end{align}
Substituting \eqref{eq:mfpt_LE_derv_n_2} in~\eqref{eq:mfpt_LE_derv_n_1} and comparing with the right-hand side of eq.~\eqref{eq:essential_edge_lemma_general}, we finally get
\begin{equation}\label{eq:mfpt_LE_conservation}
    m_{v_0 v_H}=M_{0 H}\ .
\end{equation}

We remark that eq.~\eqref{eq:mfpt_LE_conservation} generalises immediately to arbitrary pairs of vertices 
$v_I$, $v_J$ along the backbone and is not restricted to $v_0$ and $v_H$, thus {it holds that}
\begin{equation}\label{eq:mfpt_LE_conservation_all}
    m_{v_I v_J}=M_{IJ}
\end{equation}
for arbitrary $I,J$. The proof is valid for arbitrary random walkers; it implies that along the backbone of the necklace, coarse-graining according to LE preserves MFPTs. 

Since random walks on graphs with necklace structure can be coarse-grained into one-dimensional random walks, explicit MFPT formulae
can be derived, for such random walks, using 
eq.~\eqref{eq:mfpt_LE_derv_n_1}. 
In the next section, we demonstrate this method in detail for the simple random walk on c-ary trees.

In addition, we note that for graphs with necklace structure, where \eqref{eq:mfpt_LE_conservation_all} 
holds exactly, MFPTs can also be computed from  eq.~\eqref{eq:essential_edge_lemma_general}. 
Conversely, when graphs do not have necklace structure, neither eq.
~\eqref{eq:essential_edge_lemma_general} nor eq.~\eqref{eq:mfpt_LE_conservation_all} hold exactly, however, when deviations from the necklace structure are small, one may expect eq.~\eqref{eq:mfpt_LE_conservation_all} to
hold approximately.
This means that MFPTs for the coarse-grained dynamics can be used as proxies for certain MFPTs in the original dynamics. 
Importantly, the coarse-grained MFPTs can still be computed exactly via eq.~\eqref{eq:mfpt_LE_derv_n_1}, as long as the coarse-grained graph is a one-dimensional lattice. Hence the LE coarse-graining method can be used to get a reliable estimate of MFPTs in graphs with a more general structure than {that of} necklace{s}, as long as they can be coarse-grained into one-dimensional lattices.

We conclude this section by showing that for reversible random walks on non-directed graphs, 
with edge weights ${\bf e}$, such that 
$e_{ij}=e_{ji}~\forall~i,j$,  and 
transition probability matrix 
\begin{equation}
    q_{ij}=\frac{e_{ij}}{\tilde{k}_i}
    \end{equation}
with $\tilde{k}_i=\sum_j e_{ij}$,
eq.~\eqref{eq:essential_edge_lemma_general}
retrieves results from the essential edge lemma (EEL)  
\cite[Lemma 5.1]{Aldous1999}. 

Firstly, as the MFPTs between the nodes of the backbone are additive, i.e. $m_{v_I v_J}+m_{v_J v_K}=m_{v_I v_K}$ for $I<J<K$, we may also write eq.~\eqref{eq:essential_edge_lemma_general} as 
\begin{equation}
    m_{v_0 v_H} = \sum_{I=1}^{H} m_{v_{I-1} v_I}\ , 
\end{equation}
such that for each $I\geq 1$ 
\begin{equation}\label{eq:m_I-I-1}
    m_{v_{I-1} v_I}=\sum_{K=0}^{I-1}\frac{w_{I-1,K}}{w_{KI}}\frac{\Pi_K}{\pi_{v_K}}\ .    
\end{equation}
Due to the symmetry of $\bm{e}$, the equilibrium occupancy probabilities are given by 
\begin{align}
    \pi_{v_K}=\frac{\tilde{k}_{v_K}}{Z}=\frac{1}{Z}\left(e_{v_K v_{K-1}}+e_{v_{K} v_{K+1}}+\sum_{\ell=1}^{|K|-1} e_{v_K v_{K\ell}} \right)\ ,
\end{align}
where $Z=\sum_{K=0}^H\sum_{\ell=0}^{|K|-1}\tilde{k}_{v_{K\ell}}$ is the normalising factor, and we stipulate that $e_{v_0,v_{-1}}=0$ (for $K=0)$.
Similarly, the equilibrium occupation probability for the cluster $K$ is the sum of the corresponding probabilities of its vertices
\begin{align}
    \Pi_K&=\pi_{v_K}+\frac{1}{Z}\sum_{\ell =1}^{|K|-1}\sum_{m=0}^{|K|-1}e_{v_{K\ell} v_{Km}} = 
    \frac{1}{Z}\left(e_{v_{K} v_{K-1}}+e_{v_K v_{K+1}}+\sum_{m=0}^{|K|-1}e_{v_K v_{Km}}+\sum_{\ell=1}^{|K|-1}\sum_{m=0}^{|K|-1}e_{v_{K\ell} v_{Km}}\right)
    \nonumber\\
    &=
    \frac{1}{Z}\left(e_{v_{K} v_{K-1}}+e_{v_{K} v_{K+1}} + \sum_{\ell,m = 0}^{|K|-1} e_{v_{K\ell} v_{Km}}\right)\ .
\end{align} 
Moreover, the first factor in the sum of eq.~\eqref{eq:m_I-I-1} can be simplified by expanding the $w$'s according to their definition in eq.~\eqref{eq:path_weight}
\begin{equation}
    \frac{w_{I-1,K}}{w_{KI}}=\prod_{J=K+1}^{I-1}\frac{e_{v_{J},v_{J-1}} }{\tilde{k}_{v_{J}}}\prod_{J=K}^{I-1}\frac{\tilde{k}_{v_{J}}}{e_{v_{J},{v_{J+1}}} }=\frac{\tilde{k}_{v_K}}{e_{v_{I-1}v_I}} \ .
\end{equation}
This expression substituted
into eq.~\eqref{eq:m_I-I-1} produces the EEL derived in \cite[Lemma 5.1]{Aldous1999}
\begin{align}\label{eq:EEL}
    m_{v_{I-1}v_I} &= \frac{1}{e_{v_{I-1}v_I}} \sum_{K=0}^{I-1} \left(e_{v_{K} v_{K-1}}+e_{v_{K} v_{K+1}} + \sum_{\ell,m = 0}^{|K|-1} e_{v_{K\ell} v_{Km}}\right) \\ \nonumber
    &=
    1 + \frac{2}{e_{v_{I-1} v_I}} \left( \sum_{K=0}^{I-2}e_{v_K v_{K+1}} +  \sum_{K=0}^{I-1} \sum_{0 \leq \ell < m \leq |K|-1}e_{v_{K\ell} v_{Km}} \right)\ ,
\end{align}
where in the last step we have used again 
the symmetry of $\bm{e}$.

Eq.~\eqref{eq:EEL} replaces the matrix inversion 
eq.~\eqref{eq:mfpt_grounded_laplacian} by a sum over edge weights, a much less expensive operation. It provides particularly great leverage when a graph has many essential edges, the limiting case being a tree, for which every edge is essential. 

For non-directed {\em unweighted} graphs, $\bm{e}$ is replaced by an adjacency matrix $\bA$, with entries $A_{ij}\in \{1,0\}$ 
denoting presence ($1$) or absence ($0$) of links. Then the inner sum in eq.~\eqref{eq:EEL} counts the number of edges $E_K$ within subgraph $K$, and the sum $\sum_{K=0}^{I-2}e_{v_K v_{K+1}}$ counts the number of backbone edges connecting $v_0$ and $v_{I-1}$, amounting to $I-1$. Therefore, eq.~\eqref{eq:EEL} yields in this case
\begin{equation}\label{eq:EEL_0-1}
    m_{v_{I-1} v_I}= 2I-1+2\sum_{K=0}^{I-1}E_K\ .
\end{equation}
Hence for simple random walks on non-directed unweighted graphs, all that is required to compute MFPTs between the ``hanging points'' of two clusters $I$ and $J$ is the number of edges in the subgraphs $0,\dots, I$. It is important to stress however that eqs.~\eqref{eq:EEL} and ~\eqref{eq:EEL_0-1} apply only to reversible random walks, while eq. \eqref{eq:essential_edge_lemma_general} applies to general random walks.

In the next section we demonstrate how the LE coarse-graining method can be used to derive explicit {MFPT} formulae for the simple random walk on $c$-ary trees. In the subsequent sections we apply the GEEL, eq.~\eqref{eq:essential_edge_lemma_general}, to 
non-tree graphs with necklace structure and more general random walks, including irreversible random walks. 
In addition, we show the usefulness of the LE coarse-graining method
when dealing with
more general graph structures, where the EEL, even its more general formulation, GEEL, cannot be applied, e.g. as there are no essential edges in the original graphs, or calculations cannot be simplified to explicit formulae. We show that in such cases, LE coarse-graining can provide reliable MFPT approximations.

\section{Mean first passage times in \lowercase{c}-ary trees: exact results\label{sec:mfpt_tree}}
\begin{figure}
    \centering
    \includegraphics{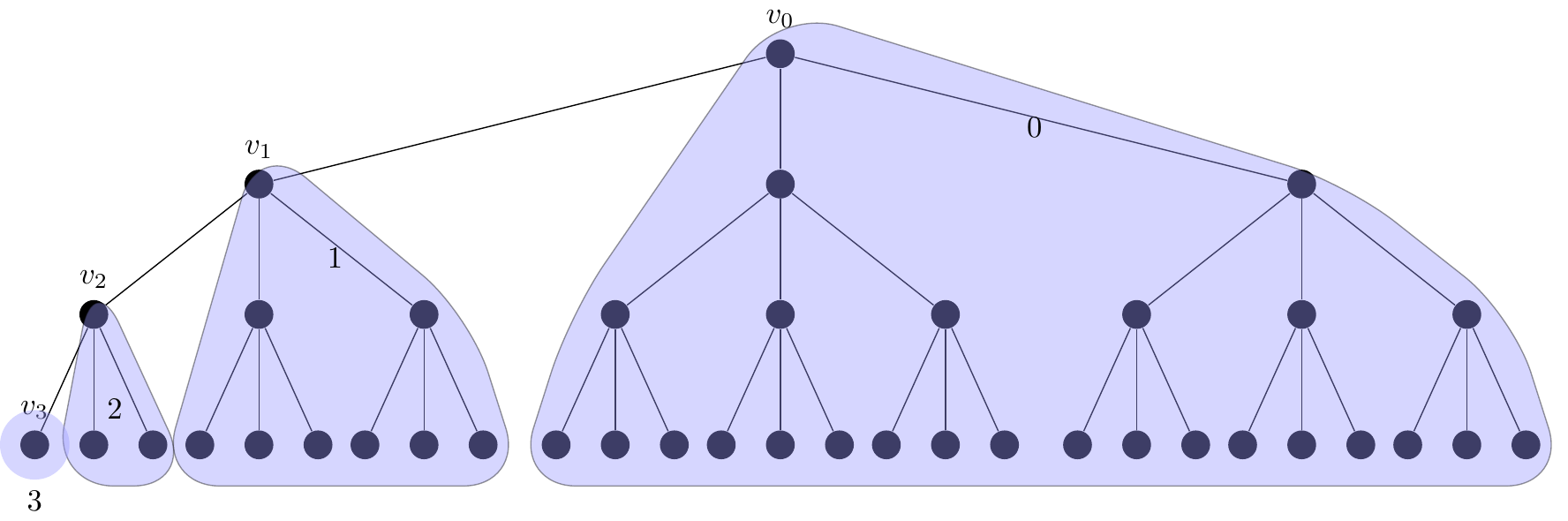}
    \caption{Ternary tree with height $H=3$. The root is $v_0$ and the target is $v_3$. Shaded areas show clusters $0,1,2,3$.}
    \label{fig:tree_clustering}
\end{figure}

In this section we 
apply the LE coarse-graining method to 
the simple random walk on 
an unweighted, non-directed $c$-ary tree of height $H$, which consists of a root with degree $c$, $H-1$ levels of descendants with degree $c+1$ (one of which corresponds to the ``upward'' edge) and a bottom level of leaves with unit degree (see  fig.~\ref{fig:tree_clustering} for an illustration). 
The transition matrix has elements
\begin{equation}
q_{ij}=\frac{A_{ij}}{k_i}\ ,
\label{eq:q_RW}
\end{equation}
where $\bm{A}$ is the adjacency matrix and 
$k_i=\sum_j A_{ij}$ is the degree of node $i$. The equilibrium occupancy probability of node $i$ is $\pi_i=k_i/(2E)$ where $E=\frac{1}{2}\sum_{i=1}^n k_i$ is the total number of links. The
MFPTs $m_{ij}$ between any two nodes $i,j$ can in principle be obtained by solving numerically the system of equations \eqref{eq:mfpt_basic_recurrence}, however, in this work we are concerned with the derivation of explicit formulae. Since the simple random walker {defined by eq.~}\eqref{eq:q_RW} is reversible, and 
every link is essential{,}
the EEL in eq.~\eqref{eq:EEL} is also applicable here (as are other methods for reversible random walkers). However, we apply here the LE coarse-graining method, with the purpose of demonstrating it on a simple example, where results are available via other methods and can be easily validated. 
Our first objective is to calculate the MFPT from the root to a target leaf, then we turn to MFPTs between 
arbitrary vertices.

\subsection{MFPT from root to leaf: exact results 
in the coarse-grained tree\label{sec:mfpt_tree_root_leaf}}
Consider a $c$-ary tree of height $H$, with root $v_0$, as shown in fig.~\ref{fig:tree_clustering}. Without loss of generality, we set the target in the first leaf, $v_H$, noting that, due to the symmetry of the tree, 
it is always possible to draw the diagram in such a way that the target is the first leaf.

The starting point of our derivation consists in
reducing the dimensionality of the problem by coarse-graining the tree according to the LE method. 
We coarse-grain the tree 
into $H+1$ subgraphs, in such a way 
that every node in the path $v_0,v_1,\dots,v_H$ (see fig.~\ref{fig:tree_clustering}) is assigned its own subgraph. We define each subgraph $I=0,\ldots, H$ as containing $v_I$, and all the vertices 
of the tree rooted in $v_I$, excluding the branches through $v_{I-1}$ and $v_{I+1}$, 
as shown by the shaded areas in fig.~\ref{fig:tree_clustering}. Note that subgraph 
$H$ contains {only the node} $v_H$. 

By comparison with fig.~\ref{fig:tree_clustering}, it is clear that $c$-ary trees belong to the class of graphs with necklace structure considered in sec.~\ref{sec:main_derivation}, hence 
the MFPT from root to leaf can be computed as the MFPT from cluster $0$ to cluster $H$ in the LE coarse-grained dynamics.

We thus define the hopping probability between clusters according to the LE definition eq.~\eqref{eq:LE_def}, where 
$\Pi_I=\sum_i C_{iI}\pi_i$. Since the $H+1$ clusters are 
sitting on
a one-dimensional lattice, the transition 
matrix $\bQ$ of the coarse-grained dynamics will be a
$(H+1)\times(H+1)$ tridiagonal matrix. Its elements are obtained from eq.~\eqref{eq:LE_def} as follows.
Writing $\sum_i C_{iI}\cdots=\sum_{i\in I}\cdots$ and $\sum_{j\in I}=\sum_j-\sum_{j\not\in I}$, we have 
\begin{equation}
Q_{II}=\frac{1}{\sum_{i\in I}k_i}\sum_{i\in I}\left(\sum_j A_{ij}-\sum_{j\not\in I} A_{ij}\right)\ .
\end{equation}
Noting that $k_i=\sum_j A_{ij}$ and that 
the number of links between 
cluster $I$ and any other cluster is 
$\sum_{i\in I, j\not\in I}A_{ij}=2$ for all $I$ (except clusters $I=0,H$ that have a single out-going edge, each), we have
\begin{equation}
    Q_{II} =\left\{
    \begin{array}{ll}
        1-\frac{1}{\sum_{i\in I}k_i} \quad & \colon I\in\{0,H\} \\
        1-\frac{2}{\sum_{i\in I}k_i} \quad & \colon 1 \leq I \leq H-1 \ ,
\end{array}
\right.
\end{equation} 
and similarly
\begin{equation}
    Q_{IJ}=\frac{1}{\sum_{i\in I}k_i}(\delta_{I,J-1}+\delta_{I,J+1})\quad \colon I,J=0\ldots H\ ,
\end{equation}
with the understanding that $\delta_{I,-1}=\delta_{I,H+1}=0$.
The sums $\sum_{i\in I}k_i$ can be derived using the following facts:
\begin{enumerate}
    \item Every subgraph $I$ consists of a top node at height $H-I$, and $c-1$ copies of the $c$-ary tree with height $H-I-1$,
    \item a $c$-ary tree with height $H$ {has}
        \begin{equation}\label{eq:c-tree_size}
        n=(c^{H+1}-1)/(c-1)
        \end{equation}
        vertices,
    \item the sum of degrees of any graph is twice the number of its edges,
    \item any tree with $n$ vertices has exactly $n-1$ edges.
\end{enumerate}
Using these observations we find
\begin{equation}
    \sum_{i\in I}k_i = \begin{cases}
    2 (|I|-1)+1=2c^{H-I}-1 \quad &\colon I\in\{0,H\}\ , \\
    2 (|I|-1)+2=2c^{H-I} \quad &\colon 1 \leq I \leq H-1 \ ,
    \end{cases}
\end{equation}
where we denoted with $|I|$ the size of subgraph $I$, i.e. the number of its vertices.
Hence, the local-equilibrium transition probabilities read
\begin{align}\label{eq:LE_tree_transition_probs}
    Q_{IJ}=&\begin{cases}
    (\delta_{J,I-1}+\delta_{J,I+1}) \frac{1}{2c^{H-I}-1} \quad &\colon I\in\{0,H\}\ , \\
    (\delta_{J,I-1}+\delta_{J,I+1}) \frac{1}{2c^{H-I}} \quad &\colon 1 \leq I \leq H-1 \ ,
    \end{cases}\\
    Q_{II} =& \begin{cases}
    \frac{2c^{H-I}-2}{2c^{H-I}-1} \quad &\colon I\in\{0,H\}\ , \\
    \frac{c^{H-I}-1}{c^{H-I}} \quad &\colon 1 \leq I \leq H-1 \ .
    \end{cases}
\end{align}

Eq.~\eqref{eq:mfpt_LE_conservation} from sec.~\ref{sec:main_derivation} {implies} that the MFPT $m_{v_0 v_H}$ from root to target in the original tree matches the MFPT $M_{0 H}$ between the first and 
last cluster in the LE coarse-grained dynamics, when the clusters are defined as above.
Thus, we can now calculate $m_{v_0 v_H}$ 
by appealing to eqs.~\eqref{eq:mfpt_LE_conservation} and \eqref{eq:mfpt_LE_derv_n_1}. 
In the latter, the fractions can be cancelled efficiently since $Q_{I,I-1}={Q_{I,I+1}}$ for $I=1,\dots,H-1$,
\begin{align}
    \frac{W_{I-1,K}}{W_{KI}}=\prod_{J=K+1}^{I-1}\frac{Q_{J,J-1}}{Q_{J,J+1}} \frac{1}{Q_{K,K+1}}=\frac{1}{Q_{K,K+1}}\ ,
\end{align}
leading us to
\begin{equation}
    M_{0 H}=  \sum_{ I=1 }^H \sum_{K=0}^{I-1} \frac{1}{Q_{K,K+1}}. 
\end{equation}
Substituting the transition probabilities from eq.~\eqref{eq:LE_tree_transition_probs}, we obtain the final result
\begin{align}\label{eq:mfpt_tree_root_leaf}
   \nonumber M_{0 H}&=2\sum_{j=1}^{H-1}j c^j + H(2c^H-1)\\
    &=H\left( \frac{2c^{H+1}}{c-1}-1  \right)-2c\frac{c^H-1}{(c-1)^2}\ .
\end{align}

The above expression matches exactly the MFPT $m_{v_0 v_H}$ in the original system (i.e. without coarse-graining), as derived in \cite[Example 5.14]{Aldous1999} by using the EEL on every edge between $v_0$ and $v_H$ and adding up the results. We note that 
thanks to the tridiagonal nature of the coarse-grained 
transition matrix $\bQ$, one could have also pursued the 
matrix inversion in eq.~\eqref{eq:mfpt_grounded_laplacian}, however computations via this route are more involved.

\subsection{MFPTs between arbitrary vertices of the \lowercase{c}-ary tree\label{sec:mfpt_tree_any_vertices}}

\begin{figure}[h]
    \centering
    \includegraphics{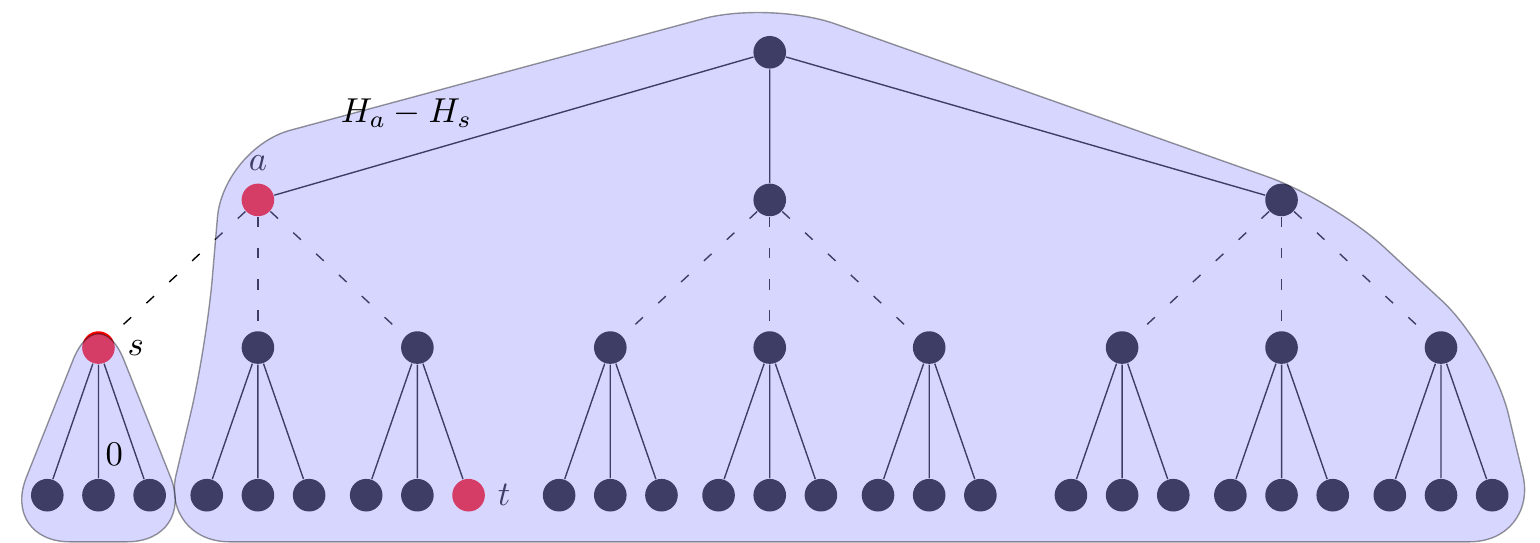}
    \caption{Ternary tree
    as shown in fig.~\ref{fig:tree_clustering}, with an example of  source $s$, target $t$ and common ancestor $a$ marked in red. Shaded areas enclose the subgraphs $0$ and ${H_a-H_s}$ defined for the first-passage process from $s$ to $a$ (intermediate clusters not shown). For this example, $H_s=1$ {and $H_t=0$}. For the first-passage process from $a$ to $s$, subgraphs are labelled in reverse order. Dashed lines indicate potentially omitted levels.}
    \label{fig:tree_clustering_general_nodes}
\end{figure}

In this section we complement the results obtained in sec.~\ref{sec:mfpt_tree_root_leaf} for the MFPT from root to leaf by deriving explicit formulae for the MFPTs between any two vertices $s$ and $t$ of a $c$-ary tree. In contrast to the previous section, here we appeal directly to eq.~\eqref{eq:essential_edge_lemma_general}, which provides an equivalent route to eq.~\eqref{eq:mfpt_LE_derv_n_1}.

Firstly, we can always permute the branches in such a way that $s$ lies on an outer branch of the diagram, as shown in fig.~\ref{fig:tree_clustering_general_nodes}.
We can then proceed by (i) finding their common ancestor $a$, which also lies on the outer branch, on the path between $s$ and the root, ii) calculating the MFPTs from the source $s$ to the ancestor $a$ and from the ancestor to the target $t$ separately, and finally (iii) adding up the results. 

{For the purpose of this section, $H$ denotes the height of the tree, not the number of clusters employed in the coarse-graining approach. For this, we}
denote by $H_s$ the height of $s$ (defined as the distance to the leaves of the tree rooted in $s$), and by $H_a$ the height of $a$. The ``upward'' MFPT $m_{s a}$ can be obtained by defining $H_a-H_s+1$ subgraphs as follows: Subgraph $0$ contains the tree rooted in $s$ excluding the branch pointing towards $a$. Subgraph ${H_a-H_s}$ is formed by the tree rooted in $a$ excluding the branch leading to $s$. For instance, in the diagram shown in fig.~\ref{fig:tree_clustering_general_nodes}, the subgraph $0$ consists of the children of $s$, and ${H_a-H_s}$ contains all branch-offs at $a$ leading away from $s$.
Following the notation introduced earlier, we identify $v_0=s$, $v_{{H_a-H_s}}=a$ and enumerate the vertices along the line connecting $s$ to $a$ as $v_1,\dots, v_{{H_a-H_s-1}}$.
Each intermediate subgraph $I$ for $I \in \{1,\dots, {H_a-H_s-1}\}$ contains the tree rooted in $v_I$
excluding both branches leading to $s$ and $a$.

In order to apply eq.~\eqref{eq:essential_edge_lemma_general}, we need to compute for each $I=1,\dots,{H_a-H_s}$ and $K=0,\dots, I-1$ the summands
\begin{equation}\label{eq:m_s_a_w_ratios}
    \frac{w_{I-1,K}}{w_{KI}}\frac{\Pi_K}{\pi_{v_K}}=
    \prod_{J=K+1}^{I-1}\frac{  q_{v_J v_{J-1}} }{q_{v_J v_{J+1} }}\frac{\Pi_K}{q_{v_K v_{K+1}} \pi_{v_K}}
\end{equation}
which follows from the definition of the weights $w$ given in eq.~\eqref{eq:path_weight}. 
To handle the product, we notice that leaves and the root and can only appear on the backbone as the source $s=v_{0}$ and the ancestor $a=v_{{H_a-H_s}}$, respectively, whereas the vertex $v_J$ with the product index $J=K+1,\dots, I-1$ runs from $v_{1}$ if $K=0$ to $v_{{H_a-H_s-1}}$ if $I={H_a-H_s}$. Therefore, the degree of $v_J$ is always $k_{v_J}=c+1$, and $q_{v_{J} v_{J+1}}=q_{v_{J} v_{J-1}}=\frac{1}{c+1}$. Consequently, the product in the above formula can be cancelled.
Inserting eq.~\eqref{eq:m_s_a_w_ratios} into eq.~\eqref{eq:essential_edge_lemma_general} then gives
\begin{align}\label{eq:m_s_a_simplified}
    m_{s a}=\sum_{I=1}^{{H_a-H_s}}\sum_{K=0}^{I-1} \prod_{J=K+1}^{I-1}\frac{  q_{v_J v_{J-1}} }{q_{v_J v_{J+1} }}\frac{\Pi_K}{q_{v_K v_{K+1}} \pi_{v_K}} =\sum_{I=1}^{{H_a-H_s}}\sum_{K=0}^{I-1} \frac{1}{q_{v_K v_{K+1}}}\frac{\Pi_K}{\pi_{v_K}}\ .
\end{align} 
Since for the simple random walker the equilibrium probabilities of the vertices are proportional of their degrees, we can expand
\begin{equation}
    \Pi_K = \sum_{\ell=0}^{|K|-1} \pi_{v_{K\ell}} = \frac{1}{Z}\sum_{ \ell=0}^{|K|-1} k_{v_{K\ell}}\ ,
\end{equation}
with normalising factor $Z$.
Similarly to the argument in sec.~\ref{sec:mfpt_tree_root_leaf}, the sum of the degrees in $K$ counts the number of edges leaving $K$, and double-counts the edges within it. Using the fact that each inner subgraph $K=1,\dots H_a-H_s$ is connected to two neighbouring subgraphs, while the outer subgraphs $K=0,\ {H_a-H_s}$ only are connected to one, each, we have
\begin{equation}\label{eq:m_sa_Pi_K}
    \Pi_K=
    \begin{cases} 
       \frac{1}{Z}{[2(|K|-1)+1]} \quad &\colon \ K\in \{0,{H_a-H_s}\} \ ,\\      
       \frac{1}{Z}{[ 2(|K|-1)+2]} \quad &\colon \ 1\leq K \leq {H_a-H_s-1}\ .
    \end{cases} 
\end{equation}
The sizes of the subgraphs $K$ are now given by
\begin{equation}\label{eq:m_s_a_cluster_size}
    |K|=\begin{cases}
        \frac{c^{H_s+1}-1}{c-1} \quad &\colon K=0\ , \\
        \frac{c^{H+1}-c^{H_a}}{c-1} \quad &\colon K={H_a-H_s}\ , \\
        c^{H_s+K} \quad &\colon 1\leq K \leq {H_a-H_s-1}
    \end{cases}
\end{equation}
with the same reasoning as in sec.~\ref{sec:mfpt_tree_root_leaf}.
Moreover, the equilibrium occupancy probabilities $\pi_{v_K}=\frac{k_{v_K}}{Z}$ as well as 
the hopping probabilities $q_{v_K v_{K+1}}=\frac{1}{k_{v_{K}}}$ can be combined into 
\begin{equation}\label{eq:m_s_a_summand}
    \frac{1}{q_{v_K v_{K+1}}}\frac{\Pi_K}{\pi_{v_K}}=\begin{cases} 
       2(|K|-1)+1 \quad &\colon \ K \in \{0,{H_a-H_s}\} \ ,\\ 
       2(|K|-1)+2 \quad &\colon \ 1\leq K \leq {H_a-H_s-1}\ .
    \end{cases}
\end{equation}
Substituting this expression together with the subgraph sizes $|K|$ {from} eq.~\eqref{eq:m_s_a_cluster_size} into eq.~\eqref{eq:m_s_a_simplified}, we arrive at the result
\begin{align}\label{eq:mfpt_tree_up}
    m_{s a}&=\sum_{I=1}^{{H_a-H_s}} \left( \sum_{K=1}^{I-1} 2|K| + 2|0|-1  \right) \\ \nonumber
    &=\sum_{I=1}^{{H_a-H_s}} \left( \sum_{K=0}^{I-1} 2c^{H_s+K} +2\frac{c^{H_s+1}-1}{c-1}-1 \right)\\ \nonumber
    &=2 \frac{c^{H_a+1}-c^{H_s+1}}{(c-1)^2}-(H_a-H_s)\frac{c+1}{c-1}\ .
\end{align}

In the opposite direction, downward from $a$ to $s$, we have to consider the $v_I$'s in reverse order
\begin{equation}
    m_{a s}=\sum_{I=1}^{{H_a-H_s}}\sum_{K=0}^{I-1}\frac{1}{q_{v_{H_a-H_s-K}, v_{H_a-H_s-K-1}}}\frac{\Pi_{H_a-H_s-K}}{\pi_{v_{H_a-H_s-K}}}\ ,
\end{equation}
into which we can again substitute eq.~\eqref{eq:m_s_a_summand} to obtain
\begin{align}\label{eq:mfpt_tree_down}
    m_{a s}&= \sum_{I=1}^{{H_a-H_s}} \left( \sum_{K=1}^{I-1} 2c^{H_a-K} + 2\frac{c^{H+1}-c^{H_a}}{c-1}-1 \right)\\ \nonumber
    &= (H_a-H_s)\left(\frac{2c^{H+1}}{c-1}-1\right)-2\frac{c^{H_a+1}-c^{H_s+1}}{(c-1)^2}\ .
\end{align}
Note that the limiting case $H_s=0$, $H_a=H$ reproduces eq.~\eqref{eq:mfpt_tree_root_leaf} for the MFPT from the root to any leaf, as it should. 

In order to obtain $m_{s t}$ we need to add $m_{s a}$ and $m_{a t}$; the latter is obtained by replacing $t$ for $s$ in eq.~\eqref{eq:mfpt_tree_down}. The final result can be written as
\begin{align}\label{eq:mfpt_tree_general}
    m_{s t} &= m_{s a}+m_{a t}\\ \nonumber
    &=2(n-1)(H_a-H_t)+2\frac{c^{H_t+1}-c^{H_s+1}}{(c-1)^2}+(H_s-H_t)+\frac{2}{c-1}(H_s-H_t)\ ,
\end{align}
where $n$ is the total number of vertices of the tree, eq.~\eqref{eq:c-tree_size}. This result is in agreement with those derived in \cite[Example 5.14]{Aldous1999}.

\section{Exact results on non-tree graphs with necklace structure\label{sec:applications_rev}}

This section applies the 
GEEL eq.~\eqref{eq:essential_edge_lemma_general}
on two further examples of the necklace type, namely the $c$-star and a concatenation of $H+1$ cliques of size $c$. 
We note that the $c$-star graph is, in fact, a $c$-ary tree of height $1$, so the results obtained here also follow trivially from those obtained in sec.~\ref{sec:mfpt_tree}, via eq.\eqref{eq:mfpt_LE_derv_n_1}.

\subsection{Star graph}
\begin{figure}[h]
    \centering
    \includegraphics{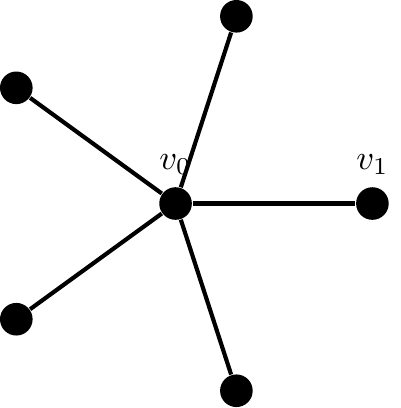}
    \caption{A star with $c=5$ vertices. The subgraph $1$ is just the vertex $v_1$; all other vertices form the subgraph $0$.\label{fig:star}}
\end{figure}

In this section we consider a star graph of $c$ vertices around a middle vertex $v_0$ as shown in fig.~\ref{fig:star}.

In analogy to what was done in sec.~\ref{sec:mfpt_tree}, we define subgraph $1$ as containing only vertex $v_1$, and subgraph $0$ as containing all other vertices.  This makes clear that the star graph
belongs to the family of necklace graphs, as required for our approach to work.

Proceeding as in the previous calculations, we apply the GEEL, eq.~\eqref{eq:essential_edge_lemma_general}, to determine $m_{v_0 v_1}$ for the simple random walker on the $c$-star. As with two clusters we have $H=1$, the GEEL in this case contains but a single summand, 
\begin{equation}\label{eq:star_GEEL}
    m_{v_0 v_1}=\sum_{I=1}^{1}\sum_{K=0}^{I-1}\frac{w_{I-1,K}}{w_{KI}}\frac{\Pi_{K}}{\pi_{v_K}}=\frac{1}{q_{v_0 v_1}}\frac{\Pi_0}{\pi_{v_0}}\ ,
\end{equation}
using that $w_{00}=1$ and $w_{01}=q_{v_0 v_1}$.
The equilibrium probabilities on the right hand side of eq.~\eqref{eq:star_GEEL} are given by $\pi_{v_0}=\frac{c}{2c}$ and $\Pi_0
=\frac{2c-1}{2c}$, because
there is one central vertex $v_0$ with degree $c$, and $c-1$ outer vertices with degree $1$ in cluster $0$. Substituting these values together with the hopping probability $q_{v_0 v_1}=\frac{1}{c}$ into eq.~\eqref{eq:star_GEEL}, we find the MFPT
\begin{equation}\label{eq:star_mfpt}
    m_{v_0 v_1}=\frac{c(2c-1)}{c}=2c-1\ .
\end{equation}

Given the simplicity of the star graph, MFPTs can be calculated explicitly, through a variety of methods. For instance, eq.~\eqref{eq:star_mfpt} could have been alternatively derived by noticing that the random walker steps to any outer node with uniform probability at every second step, and from a leaf back to the central vertex at every other step \cite{Aldous1999}. On the other hand, as noted above, the star graph is a $c$-ary tree with height $1$, therefore it adheres to eq.~\eqref{eq:mfpt_tree_root_leaf}. Finally, the matrix inversion eq.~\eqref{eq:mfpt_grounded_laplacian} can also be performed directly \cite{Bartolucci2021}.

\subsection{Cliques on a necklace\label{sec:cliques}}

\begin{figure}[h]
    \centering
    \includegraphics{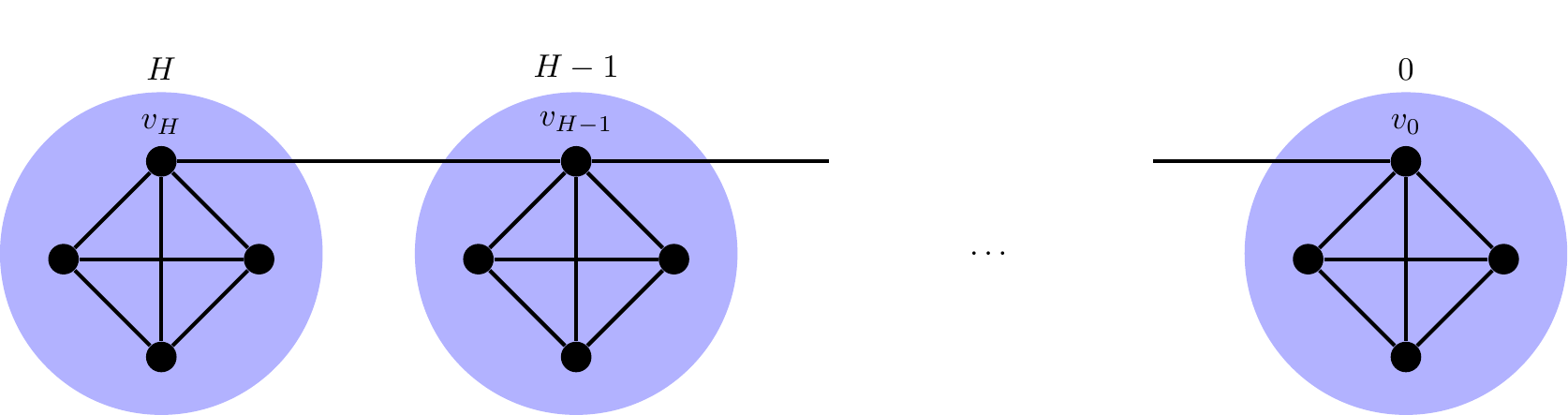}
    \caption{A necklace with $H+1$ clusters, each of which is a clique with $c=4$ vertices.\label{fig:clique_necklace}
    }
\end{figure}

Consider $H+1$ cliques, i.e. complete subgraphs, of size $c$, arranged in a chain passing through the hanging points $v_0,\dots,{v_H}$; an example for $c=4$ is shown in fig.~\ref{fig:clique_necklace}. 
As this graph has necklace structure, we will again apply the GEEL, eq.~\eqref{eq:essential_edge_lemma_general}, to determine $m_{v_0 v_H}$ for the simple random walker on this graph.

The transition probabilities between the nodes of the chain for the simple random walker are given by
\begin{equation}
    q_{v_I v_J}=\begin{cases}
        (\delta_{I,J-1}+\delta_{I,J+1}) \frac{1}{c} \quad &\colon I\in \{0,H\} \ , \\
        (\delta_{I,J-1}+\delta_{I,J+1}) \frac{1}{c+1} \quad & \colon 1\leq I \leq H-1\ ,
    \end{cases}
\end{equation}
as the degrees $k_{v_I}$ are either $c$ (first case) or $c+1$ (other cases). As in eq.~\eqref{eq:m_s_a_simplified} for the $c$-ary tree, this implies that the summands in eq.~\eqref{eq:essential_edge_lemma_general} simplify,
\begin{align}\label{eq:cliques_GEEL}
    m_{v_0 v_H} = \sum_{I=1}^{H}\sum_{K=0}^{I-1}\frac{w_{I-1,K}}{w_{KI}}\frac{\Pi_K}{\pi_{v_K}}= \sum_{I=1}^{H}\sum_{K=0}^{I-1}
    \frac{\Pi_K}{q_{v_K v_{K+1}} \pi_{v_K}} \ .
\end{align}
To determine the equilibrium probabilities, we notice that within each subgraph there are $c-1$ vertices $v_{Ki}$ ($i\neq 0$) with degree $k_{v_{Ki}}=c-1$, while the hanging vertex $v_K$ has degree $k_{v_K}=c+1$ if $K=1,\dots,H-1$ or ${k_{v_{K}}}=c$ if $K\in \{0,H\}$. Hence the stationary probability ratios amount to
\begin{equation}
    \frac{\pi_{v_K}}{\Pi_K}=\frac{k_{v_K}}{k_{v_K}+(c-1)^2}=\begin{cases}
        \frac{c}{c+(c-1)^2} \quad &\colon K=0,H\ , \\
        \frac{c+1}{c+1+(c-1)^2} \quad &\colon 1\leq K\leq H-1\ .
    \end{cases}
\end{equation}
Given that the random walker is the simple random walker, we can now write the summands in eq.~\eqref{eq:cliques_GEEL} as
\begin{equation}
     \frac{\Pi_K}{q_{v_{K} v_{K+1}}\pi_{v_{K}}} =k_{v_I}+(c-1)^2\ ,
\end{equation}
which allows us to conclude
\begin{align}\label{eq:mfpt_cliques}
 \nonumber   m_{v_0 v_H}&=\sum_{I=1}^H \left( \sum_{K=1}^{I-1} \left( c+1+(c-1)^2\right) +c+(c-1)^2 \right) \\
    &=\frac{c(c-1)H(H+1)}{2}+H^2\ .
\end{align}

For a simple path of length $H$ we have $c=1$, hence reproducing the well-known $m_{v_0 v_H}=H^2$ \cite{Lovasz1994}. For later reference, we also note that for $H=1$ and arbitrary $c$, we find $m_{v_0 v_1}=c(c-1)+1$.

\section{Applications to irreversible random walks\label{sec:applications_non_rev}}

In this section, we show that 
the GEEL, eq.~\eqref{eq:essential_edge_lemma_general}, 
can also be applied to irreversible random walks (on necklace graphs), where
the classical EEL{, eq.}~\eqref{eq:EEL}, is 
not applicable. 
To this purpose, we consider below two simple examples where results can be validated by direct computations.

\subsection{Example 1 -- irreversible random walk}
\label{sec:applications_non_rev_ex1}
\begin{figure}[h]
    \centering
    \includegraphics{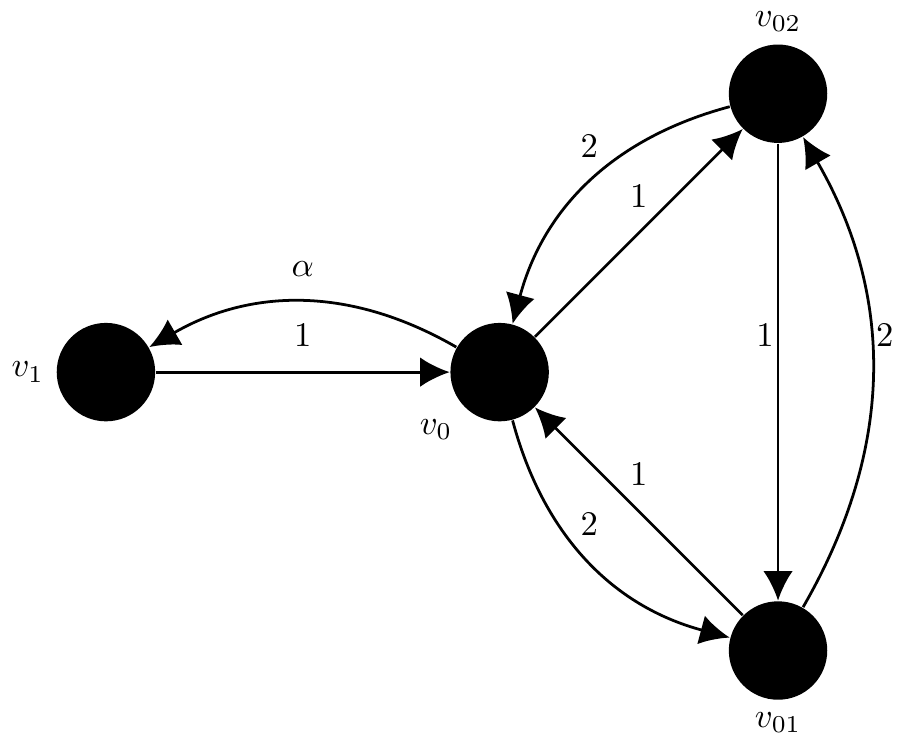}
    
    \caption{A weighted graph with asymmetrical edge weights that define an irreversible Markov chain on its vertices. }
    \label{fig:irreversible_MC_1}
\end{figure}

Consider the Markov chain in {fig.~}\ref{fig:irreversible_MC_1}, 
representing a random walk  
with transition matrix 
\begin{equation}\label{eq:irr1_q}
    \bq=\begin{pmatrix}
        0 & \frac{2}{3+\alpha} & \frac{1}{3+\alpha} & \frac{\alpha}{3+\alpha}   \\
        \frac{1}{3} & 0 & \frac{2}{3} & 0 \\ 
        \frac{2}{3} & \frac{1}{3} & 0 & 0 \\
        1 & 0 & 0 & 0 \\ 
    \end{pmatrix}\ ,
\end{equation}
where the order of the states has been chosen as $v_0,v_{01},v_{02},v_1$.
For the avoidance of doubt, in fig.~\ref{fig:irreversible_MC_1}
we have labelled edges using unnormalised weights $e_{ij}$.
Given the stationary probability vector
\begin{equation}\label{eq:irr1_pi}
    \bm{\pi}^T=\frac{1}{9+2\alpha}(3+\alpha,3,3,\alpha)
\end{equation}
we can confirm that detailed balance with the transition probabilities above is not satisfied, for instance for the nodes $v_{02}$ and $v_{0}$, where the probability 
fluxes
\begin{equation}
    \pi_{v_0}q_{v_0 v_{02}}=\frac{3+\alpha}{9+2\alpha}\frac{1}{3+\alpha}=\frac{1}{9+2\alpha}
\end{equation}
and 
\begin{equation}
    \pi_{v_{02}}q_{v_{02} v_0}=\frac{3}{9+2\alpha}\frac{2}{3}=\frac{2}{9+2\alpha}
\end{equation}
do not equate. As a consequence, the EEL, eq.~\eqref{eq:EEL}, is not applicable. 
On the other hand, we can calculate MFPTs via the GEEL, eq.~\eqref{eq:essential_edge_lemma_general}, which does not rely on dynamical reversibility. With the subgraphs for eq.~\eqref{eq:essential_edge_lemma_general} defined as $0=\{v_0,v_{01},v_{02}\}$ and $1=\{v_1\}$, we have $H=1$, which means that as in eq.~\eqref{eq:star_GEEL} there is a single summand in the GEEL. Into this we substitute the details of this example, eqs.~\eqref{eq:irr1_q} and \eqref{eq:irr1_pi}
\begin{equation}
    m_{v_0 v_1}=\frac{1}{q_{01}}\frac{\Pi_0}{\pi_{v_0}}=\frac{9+\alpha}{\alpha}\ .
    \label{eq:irr1_LE}
\end{equation}

This result is easily validated by computing MFPTs directly, using 
eq.~\eqref{eq:mfpt_grounded_laplacian}. This requires the first row sum of the inverse
\begin{align}
    \left(\Eye_{3}-\widehat{\bq}_{v_1} \right)^{-1} = 
    \frac{1}{\alpha}\begin{pmatrix}
        3+\alpha & 3 & 3 \\
        3+\alpha & \frac{9\alpha+21}{7} & \frac{6\alpha+21}{7} \\ 
        3+\alpha & \frac{3\alpha+21}{7} & \frac{9\alpha+21}{7}\\
    \end{pmatrix},
\end{align}
which gives the same result as eq.~\eqref{eq:irr1_LE}.

\subsection{Example 2 -- irreversible random walk\label{sec:applications_non_rev_ex2}}
\begin{figure}
    \centering
    \includegraphics{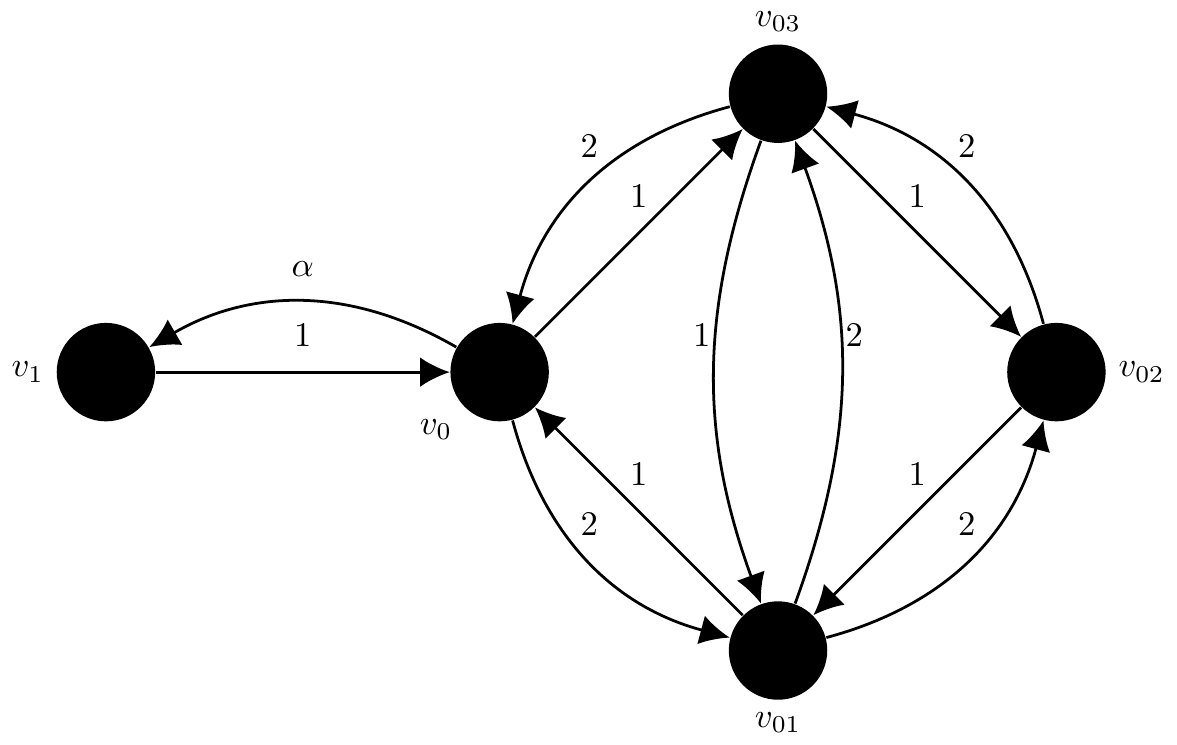}
    \caption{A weighted graph extending the example in fig.~\ref{fig:irreversible_MC_1}. }
    \label{fig:irreversible_MC_2}
\end{figure}
As a second example, 
we consider 
the graph in fig.~\ref{fig:irreversible_MC_2}. The edge weights define a random walker with transition matrix (in the order $v_0,v_{01},v_{02}, v_{03}, v_1$)
\begin{equation}
    \bm{q}=\begin{pmatrix}
        0 & \frac{2}{3+\alpha} & 0 & \frac{1}{3+\alpha} & \frac{\alpha}{3+\alpha} \\
        \frac{1}{5} & 0 & \frac{2}{5} & \frac{2}{5} & 0 \\ 
        0 & \frac{1}{3} & 0 & \frac{2}{3} & 0 \\
        \frac{1}{2} & \frac{1}{4} & \frac{1}{4} & 0 & 0\\
        1 & 0 & 0 & 0 & 0\\ 
    \end{pmatrix}
\end{equation}
which has the stationary distribution
 \begin{equation}
     \bm{\pi}^T=\frac{1}{20\alpha+141}(30+10\alpha,40,27,44,10\alpha)\ .
 \end{equation}
Defining the subgraphs $0=\{v_0,v_{01},v_{02},v_{03}\}$, $1=\{v_1\}$ for eq.~\eqref{eq:essential_edge_lemma_general}, we proceed as in the previous section to find
the MFPT
 \begin{equation}
     m_{v_0 v_1} = \frac{3+\alpha}{\alpha} \frac{\Pi_0}{\pi_{v_0}}=\frac{3+\alpha}{\alpha} \frac{10\alpha+141}{10\alpha+30}=1+\frac{141}{10\alpha}\ .
 \end{equation}
This result again is easily validated using eq.~\eqref{eq:mfpt_grounded_laplacian} and the first row sum of the inverse
\begin{align}
    \left(\Eye_{3}-\widehat{\bq}_{v_1} \right)^{-1} = 
    \frac{1}{\alpha}\begin{pmatrix}
        3+\alpha & 4 & \frac{27}{10} & \frac{22}{5}  \\
        3+\alpha & \frac{5\alpha+12}{3} & \frac{10\alpha+27}{10} & \frac{20\alpha+22}{15} \\ 
        3+\alpha & \frac{2\alpha+12}{3} & \frac{7\alpha+27}{10} & \frac{130\alpha+330}{75}\\
    \end{pmatrix}\ .
\end{align}

\section{Using LE as an approximation}\label{sec:LE_approx_for_mfpts}

In sec.~\ref{sec:main_derivation} 
we derived an exact equation for MFPTs in graphs with necklace structure, eq.~\eqref{eq:essential_edge_lemma_general}, and have proven its equivalence with eq.~\eqref{eq:mfpt_LE_derv_n_1}. We have 
shown that both equations lead to explicit {MFPT} formulae when steady-state probabilities are known.  
In this section, we show that
eq.~\eqref{eq:mfpt_LE_derv_n_1} can be used to get reliable 
approximations when the system under study deviates from the ideal setting laid out in the previous sections. In particular, we consider below two examples of graphs without necklace structure.
In the first example, considered in sec.~\ref{sec:leaks_exact_pi}, graph symmetries allow us to easily calculate the equilibrium probabilities $\bpi$. Conversely, in the second example, considered in sec.~\ref{sec:leaks_approx}, we forego such symmetries and employ a crude approximation for $\bpi$. The agreement is very reasonable in both cases,
as long as the subgraphs contain most of the total edge weight within themselves.

\subsection{Leaks into target subgraph\label{sec:leaks_exact_pi}}
\begin{figure}[h]
    \centering
    \includegraphics{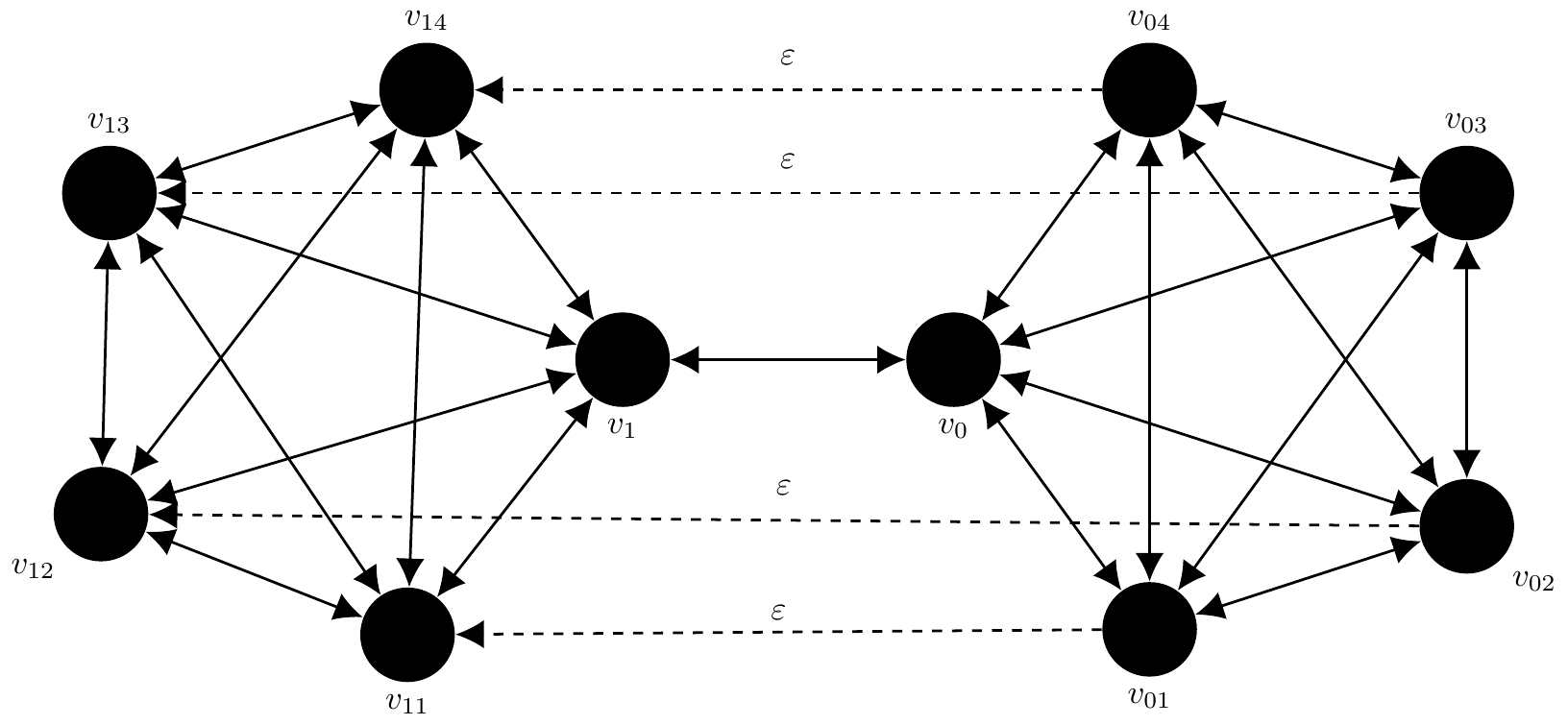}
    \caption{Two cliques $0=\{v_0,v_{01},v_{02},v_{03},v_{04}$\} and $1=\{v_1,v_{11},v_{12},v_{13},v_{14}\}$ of size $c=5$, connected to the backbone $v_0,v_1$. Additional directed edges $(v_{0i},v_{1i})$ with weight $\veps$ (``leaks'', dashed) interconnect the cliques; for every edge with weight $\veps$ the reverse edge has weight $\delta$. Unlabelled edges carry unit weight.}
    \label{fig:leaks_exact_pi}
\end{figure}

In this section we consider graphs with a structure that deviates from the necklace in that there are multiple links between two clusters. 
To this end we consider the graph
in fig.~\ref{fig:leaks_exact_pi}, consisting of 
two cliques of equal size $c$, that we will regard as cluster $0$ and $1$, respectively. 
We assume that there is a dominant link between the two clusters and all of the other inter-cluster links are weak.  
In particular, we assume that node $v_0$ in cluster $0$ is connected to node $v_1$ in cluster $1$ by two directed 
edges $(v_0,v_1)$ and $(v_1,v_0)$ with unit weight, that 
we will refer to, with a slight abuse of terminology,  
as the ``backbone'' edges. In addition, 
each vertex $v_{0i}$ in cluster $0$ is paired with a unique vertex $v_{1i}$ in cluster $1$ by 
two directed edges, 
$(v_{0i}, v_{1i})$ and $(v_{1i},v_{0i})$, 
with weight $\veps$ and 
$\delta$, respectively, which are assumed small.
As these edges ``bypass'' the backbone edges,
we refer to them as \emph{leaks} between the two clusters.

Since now the edge $(v_0,v_1)$ is inessential (removing it does not leave the graph disconnected), applying eq.~\eqref{eq:mfpt_tree_formula} becomes a formidable combinatorial task: 
the simplifications arising in eq.~\eqref{eq:essential_edge_lemma_general} do not apply here as they are only valid for necklace structures.  
However, we can consider the coarse-grained graph, with clusters defined as 
$0=\{v_0,v_{01},v_{02},v_{03},v_{04}$\} and $1=\{v_1,v_{11},v_{12},v_{13},v_{14}\}$, 
which does have an essential edge, and treat the MFPT $M_{0 1}$ in the coarse-grained dynamics as an approximation to $m_{v_0 v_1}$,
assuming that violations of 
eq.~\eqref{eq:mfpt_LE_conservation} are small if the leak weights $\veps$ and $\delta$ are small. 
Thanks to the necklace structure of the coarse-grained graph, eq.~\eqref{eq:mfpt_LE_derv_n_1} holds exactly and gives
\begin{equation}\label{eq:M01}
    M_{0 1}= \frac{W_{0,0}}{W_{0 1}}=\frac{1}{Q_{01}}
\end{equation}
where we have used the definition for $W_{IK}$ in eq.~\eqref{eq:path_weight}. In order to evaluate $Q_{01}$, as given from eq.~\eqref{eq:LE_def}, we need to calculate the equilibrium probability vector $\bpi$.

It is clear, by the symmetry of the graph in  fig.~\ref{fig:leaks_exact_pi}, that there are four classes of nodes, represented by $v_{0}, v_{01}, v_{1}, v_{11}$, and that the nodes in each class share the same equilibrium probability. To compute $\bpi$, it is then convenient to use a reduced representation of the dynamics, in terms of classes (rather than nodes). To this purpose, we note that the hopping probability between nodes of different classes is given by the $4\times 4$ matrix
\begin{align}\label{eq:q_red}
    \bm{\mathfrak{q}}^\textup{red}=\begin{pmatrix}
        0 & \frac{1}{c} & \frac{1}{c} & 0 \\
        \frac{1}{c-1+\veps} & 0 & 0 & \frac{\veps}{c-1+\veps} \\
        \frac{1}{c} & 0 & 0 & \frac{1}{c} \\
        0 & \frac{\delta}{c-1+\delta} & \frac{1}{c-1+\delta} & 0 
    \end{pmatrix} \ .
\end{align}
This matrix is not row-normalised as it only shows the hopping probabilities 
between the representatives $v_0,v_{01}, v_{1}, v_{11}$. 
Additionally, between nodes of the same class the hopping probabilities amount to
\begin{align}
    q_{v_{0i} v_{0j}}&=\frac{1}{c-1+\veps} \\ 
    q_{v_{1i} v_{1j}}&=\frac{1}{c-1+\delta} \ . 
\end{align}
Writing the eigenvector equation $\bpi^T=\bpi^T {\bf q}$ for $\pi_{v_0}$, for instance, we then find
\begin{align}
    \pi_{v_0} &= \sum_{j=0}^{c-1} \pi_{v_{0j}} q_{ v_{0j} v_{0}}  +\sum_{j=0}^{c-1} \pi_{v_{1j}} q_{ v_{1j} v_{0}} \\ \nonumber
    &= \pi_{v_0}q_{v_0 v_0} + (c-1) \pi_{v_{01}} q_{ v_{01} v_{0}} +\pi_{v_1} q_{v_1 v_0} +(c-1) \pi_{v_{11}} q_{ v_{11} v_{0}}\ .
\end{align}
In the second equality, we used that there are $c-1$ nodes in the symmetry class represented by $v_{01}$ and $v_{11}$, each. With the hopping probabilities between classes given in eq.~\eqref{eq:q_red}, $\pi_{v_0}$ simplifies to
\begin{equation}
    \pi_{v_0}=\frac{c-1}{c-1+\veps}\pi_{v_{01}}+\frac{1}{c}\pi_{v_1}\ .
\end{equation}
The remaining elements of $\bpi$ for each symmetry class can be expanded in the same way leading to the system of equations
\begin{align}
    \pi_{v_{01}} &= \frac{1}{c}\pi_{v_0} + \frac{c-2}{c-1+\veps}\pi_{v_{01}} + \frac{\delta}{c-1+\delta}\pi_{v_{11}}\ , \\ 
    \pi_{v_{1}} &= \frac{1}{c}\pi_{v_0} + \frac{c-1}{c-1+\delta}\pi_{v_{11}}\ , \\ 
    \pi_{v_{11}} &= \frac{\veps}{c-1+\veps}\pi_{v_{01}} + \frac{1}{c}\pi_{v_1} + \frac{c-2}{c-1+\delta}\pi_{v_{11}}\ . 
\end{align}
Solving the above (reduced) set of equations gives
\begin{equation}\label{eq:pi_red}
    \begin{pmatrix}
         \pi_{v_{0}} \\
         \pi_{v_{01}} \\
         \pi_{v_{1}} \\
         \pi_{v_{11}}
    \end{pmatrix} = \frac{1}{Z}
    \begin{pmatrix}
        c(1+c\delta+\veps) \\(1+\delta+c\delta)(c-1+\veps)\\ c(1+\delta+c\veps)\\ (c-1+\delta)(1+\veps+c\veps) ) 
    \end{pmatrix}\ ,
\end{equation}
with normalising constant $Z$. 

We can now calculate the LE hopping probabilities, eq.~\eqref{eq:LE_def}, again grouping nodes by their symmetry classes
\begin{align}
    Q_{01}=\frac{1}{\Pi_0}\sum_{i,j=0}^{c-1}q_{v_{0i} v_{1j}}\pi_{v_{0i}}=\frac{1}{\Pi_0} \left( q_{v_0 v_1}\pi_{v_0} + (c-1)q_{ v_{01}v_{11} }\pi_{v_{01}} \right) \ .
\end{align}
Substituting the entries for $\bpi$ obtained in eq.~\eqref{eq:pi_red} and the hopping probabilities from eq.~\eqref{eq:q_red}, we find the expression
\begin{equation}
    Q_{01}
    =\frac{1+c\delta+\veps+\veps(c-1)(1+\delta+c\delta)}{c(1+\veps+c\delta)+(c-1)(c-1+\veps)(1+\delta+c\delta)}\ ,
\end{equation}
which, due to eq.~\eqref{eq:M01}, is the reciprocal of the MFPT 
\begin{equation}\label{eq:mfpt_leaks_exact_pi_approx}
    M_{01}=\frac{1}{Q_{01}}=\frac{c(1+\veps+c\delta)+(c-1)(c-1+\veps)(1+\delta+c\delta)}{1+c\delta+\veps+\veps(c-1)(1+\delta+c\delta)} \ .
\end{equation}

\begin{figure}[h]
    \centering
    \includegraphics{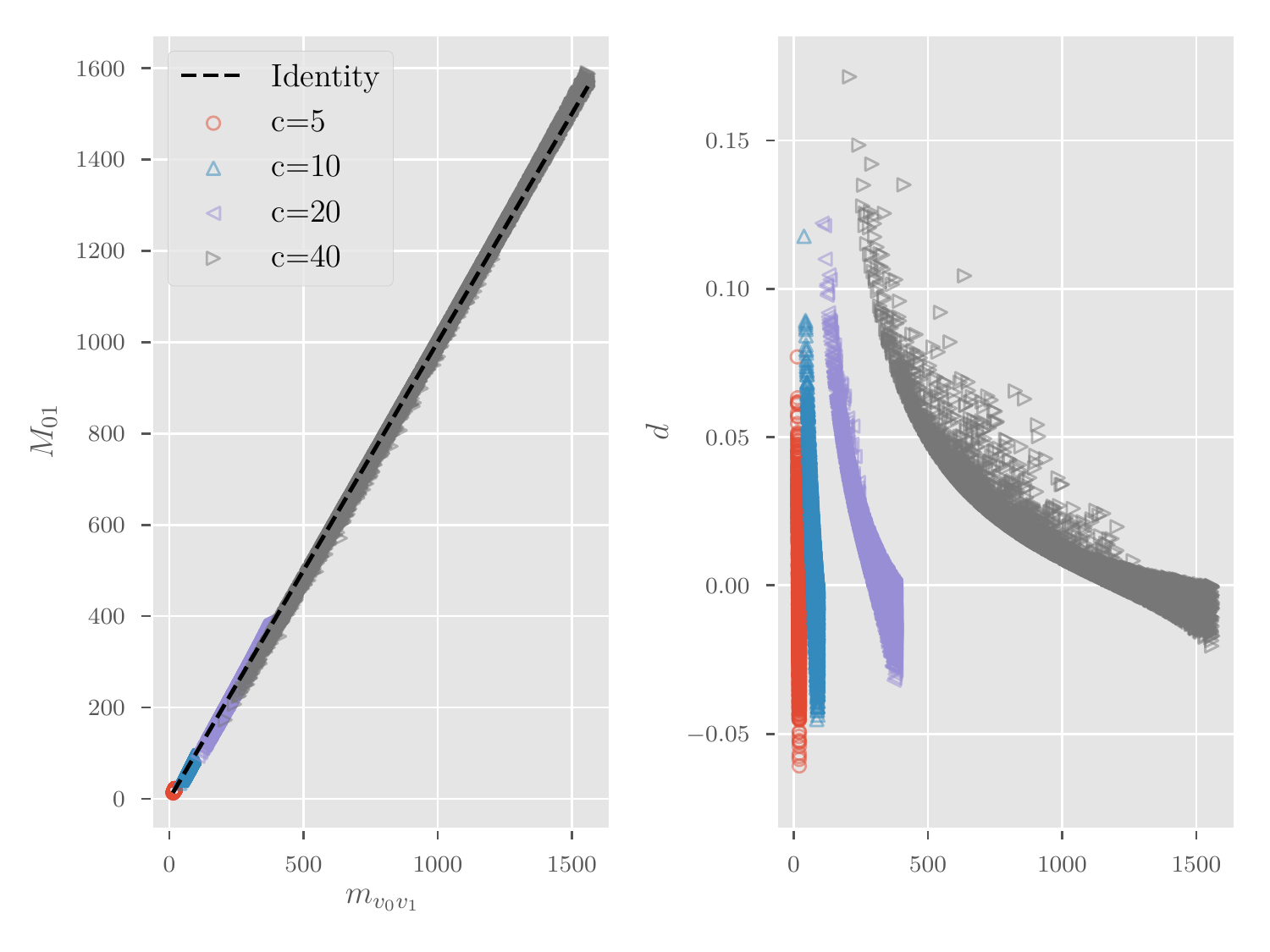}
    \caption{
    Left: Exact MFPTs $m_{v_0 v_1}$ according to eq.~\eqref{eq:mfpt_grounded_laplacian} vs approximate results $M_{01}$ as per eq.~\eqref{eq:mfpt_leaks_exact_pi_approx}, for the random walker with leaks of sec.~\ref{sec:leaks_exact_pi}. Right: Relative deviation, see eq.~\eqref{eq:rel_dv_def}.
    Each data-point represents a pair of $\veps$ and $\delta$ drawn independently with density $f_{1/50}$ as in eq.~\eqref{eq:pdf_exp}. Colours indicate the clique size $c$. For each $c$, $5000$ samples for $\veps$ and $\delta$ were drawn.
    }
    \label{fig:leaks_exact_pi_exact_vs_LE_ana}
\end{figure}

Fig.~\ref{fig:leaks_exact_pi_exact_vs_LE_ana} compares the above approximation to the true value of $m_{v_0 v_1}$, computed solving eq.~\eqref{eq:mfpt_grounded_laplacian} numerically for different values of $c$ and different values of $\veps$ and $\delta$, i.i.d. sampled from the exponential probability density
\begin{equation}\label{eq:pdf_exp}
    f_{m}(x) = \frac{1}{m} e^{-x/m}, \quad x \geq 0 
\end{equation}
with mean $m=1/50$.
The panel on the right of the figure shows that the relative deviations
\begin{equation}\label{eq:rel_dv_def}
    d=\frac{m_{v_0 v_1}- M_{0 1}}{m_{v_0 v_1}}
\end{equation}
reach higher values with increasing $c$. As the total mean leakage amounts to $c/50$ (in any direction, so to and from $0$, respectively), this means that the approximation becomes worse as more weight is accumulated on the non-dominating edges between the clusters. This is as expected, for the higher the total leakage, the further the graph is from having necklace structure, and the less precise the approximation eq.~\eqref{eq:mfpt_LE_conservation_all}. 

\subsection{Random leaks into target subgraph\label{sec:leaks_approx}}

\begin{figure}[h]
    \centering
    \includegraphics{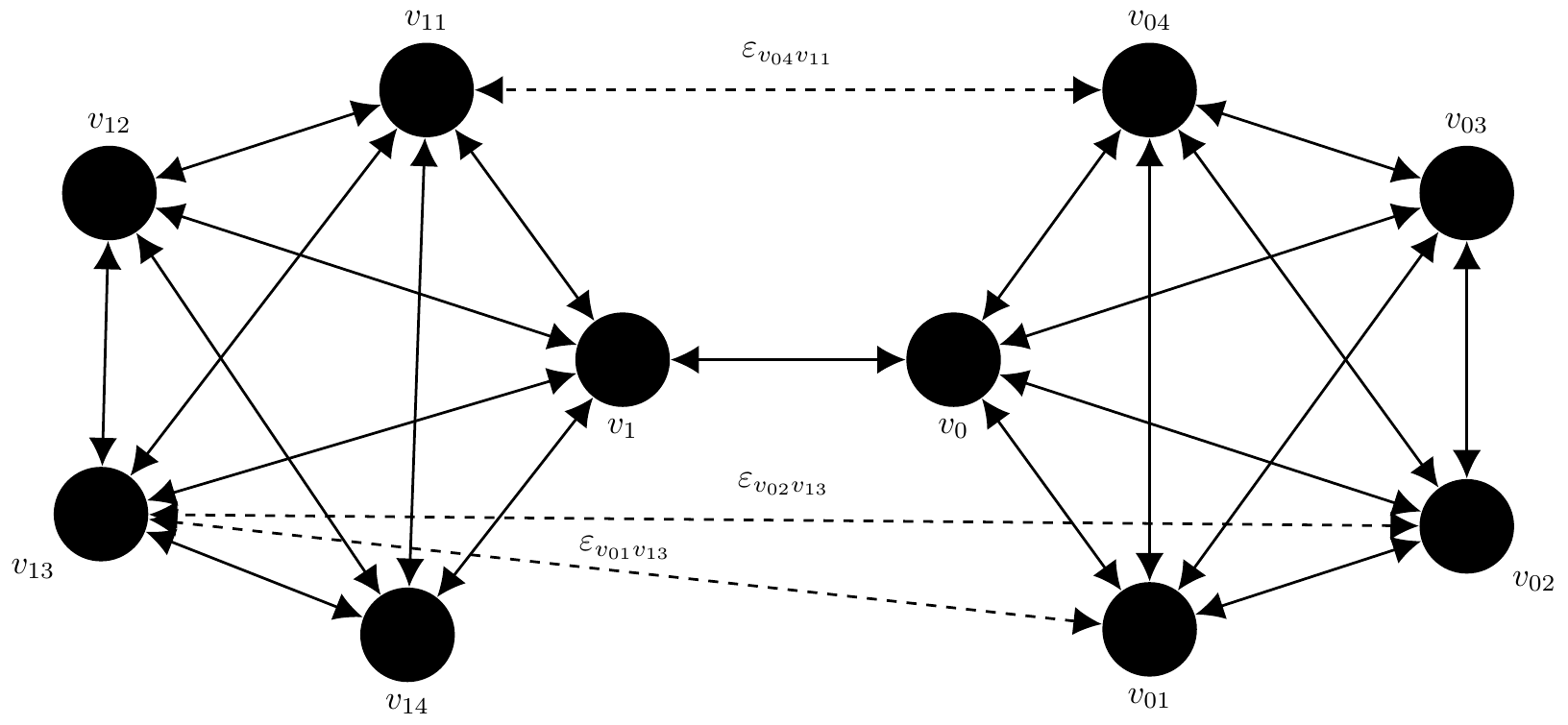}
    \caption{Two cliques $0=\{v_0,v_{01},v_{02},v_{03},v_{04}$\} and $1=\{v_1,v_{11},v_{12},v_{13},v_{14}\}$ of size $c=5$, connected to the backbone $v_0,v_1$. Some additional edges $(v_{0i},v_{1j})$ with weight $\veps_{v_{0i} v_{0j}}$ (``leaks'', dashed) interconnect the cliques. Unlabelled edges have unit weight.}
    \label{fig:leaks_approx}
\end{figure}

Here we generalise the situation of the previous sec.~\ref{sec:leaks_exact_pi} to one where $\bpi$ is less accessible. To this end we consider again a 
graph formed by two cliques of size $c$ linked by the backbone edges $(v_0,v_1)$ and $(v_1,v_0)$ with unit weight,
as in fig.~\ref{fig:leaks_approx}, where 
a fixed number, say $k$, of
additional weighted edges 
``bypass'' the backbone $(v_0,v_1)$; for simplicity of the exposition, we assume that these edges do not involve $v_0$ and $v_1$. 
As in sec.~\ref{sec:leaks_exact_pi}, we refer to them as \emph{leaks} between the two clusters
and assign them weights $\veps_{v_{0i}v_{1j}}$ in order to treat them perturbatively. 

Again, we treat the MFPT $M_{0 1}$ of the coarse-grained graph as an approximation to $m_{v_0 v_1}$, in line with eq.~\eqref{eq:mfpt_LE_conservation}, but in contrast to sec.~\ref{sec:leaks_exact_pi} we crudely approximate $\bpi$ by its value for $\veps_{v_{0i}v_{1j}}=0$, given in sec.~\ref{sec:cliques}. Considering two subgraphs, we can appeal directly to eq.~\eqref{eq:M01} to compute $M_{01}$ as in the previous example.

For the approximate coarse-grained hopping probability $Q_{01}$ we apply the definition in eq.~\eqref{eq:LE_def}
\begin{equation}
    Q_{01} = \frac{1}{\Pi_0} \left(\pi_{v_0}\frac{1}{c}+ \sum_{i=1}^{c-1} \sum_{j=1}^{c-1} \pi_{v_{0i}} \frac{\veps_{v_{0i}v_{1j}}}{c-1+\sum_{m=1}^{c-1} \veps_{v_{0i}v_{1m}} } \right)\ .
\end{equation}
Upon introducing the short-hand notation $\veps_{v_{0i}} = \sum_{m=1}^{c-1}\veps_{v_{0i}v_{0m}}$ for the total leakage at $v_{0i}$, and using the approximations $\pi_{v_0}\approx \frac{c}{Z}$ and $\pi_{v_{0i}}\approx\frac{c-1}{Z}$, we obtain
\begin{equation}\label{eq:mfpt_Q01_leaks_approx}
    Q_{0 1} \approx \frac{1}{c+(c-1)^2}\left(1+ (c-1)\sum_{i=1}^{c-1} \frac{\veps_{v_{0i}}}{c-1+\veps_{v_{0i}}} \right)\ .
\end{equation}
Finally, using eq.~\eqref{eq:M01}, we obtain 
\begin{equation}\label{eq:mfpt_leaks_approx}
    M_{0 1}=\frac{1}{Q_{01}} \approx \frac{c(c-1)+1}{1+(c-1) \sum_{i=1}^{c-1}\frac{\veps_{v_{0i}}}{c-1+\veps_{v_{0i}}}  } \ .
\end{equation}

\begin{figure}
    \centering
    \includegraphics{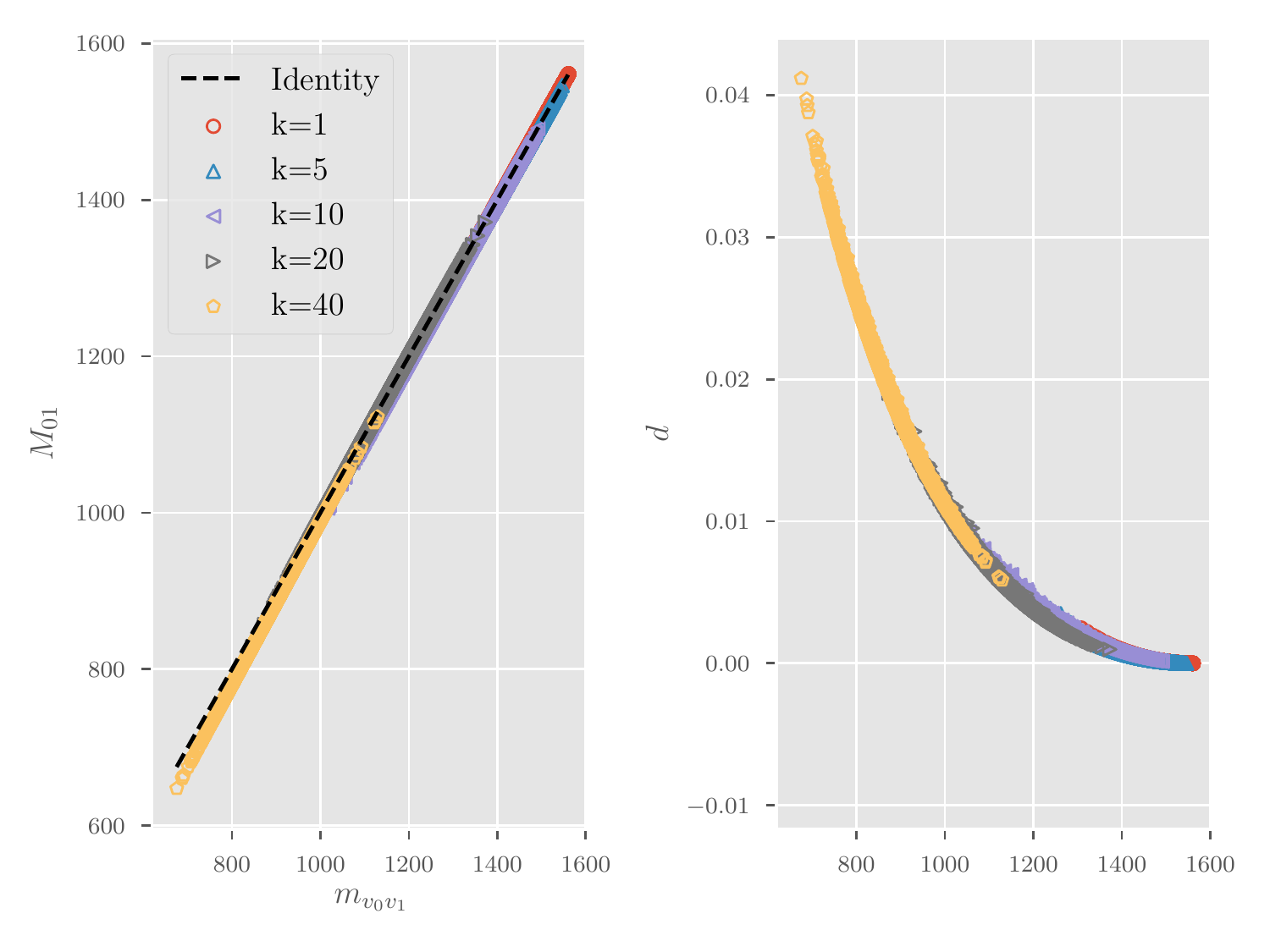}
    \caption{Left: Exact MFPTs $m_{v_0 v_1}$ according to eq.~\eqref{eq:mfpt_grounded_laplacian} vs approximate results for $M_{01}$ as per eq.~\eqref{eq:mfpt_leaks_approx}, for the random walker with leaks of sec.~\ref{sec:leaks_approx} with $c=40$. Right: Relative deviation, see eq.~\eqref{eq:rel_dv_def}.
    $k$ is the total number of leaks bypassing the backbone, drawn without replacement from all possible pairings of vertices $(v_{0i},v_{1j})$; the values for the $\veps_{v_{0i} v_{0j}}$'s are drawn independently with density $f_{1/50}$ as in eq.~\eqref{eq:pdf_exp}. For each value of $k$, $10,000$ realisations of leaks and $\veps_{v_{0i} v_{0j}}$'s were sampled.}
    \label{fig:leaks_scatter}
\end{figure}

Fig.~\ref{fig:leaks_scatter} compares the 
(approximate) value of the clustered MFPT $M_{01}$
to the true value of the original MFPT $m_{v_0v_1}$, obtained by solving eq.~\eqref{eq:mfpt_grounded_laplacian} numerically for different numbers of leaks $k$, and different values of $\veps$'s, i.i.d. sampled from the probability density $f_{\frac{1}{50}}$ as in eq.~\eqref{eq:pdf_exp}. The clique size is fixed to $c=40$.

Firstly, we notice that the relative deviation, eq.~\eqref{eq:rel_dv_def}, in the right panel of the figure increases with $k$, as the total leakage increases. This is again as expected, since higher leakage moves the graph further from having necklace structure. However, even for the largest tested value of $k$ the prediction via $M_{0 1}$ is in good agreement with the true value. Secondly, $M_{0 1}$ systematically underestimates the true MFPT; this is again reasonable to expect, for the leaks provide additional pathways into the target subgraph $1$, wherein the target $v_1$ can be reached from any node. With $k=0$, only the backbone exists and we recover the necklace-case.

\clearpage
\section{Conclusions\label{sec:conclusion}}

In this paper, we explored the behaviour of mean first-passage times of random walkers on graphs under the ``local-equilibrium'' (LE) approximation. We show that for graphs resembling a necklace -- with subgraphs arranged linearly and hanging via a single vertex from a one-dimensional chain -- the end-to-end MFPT of the ``natural'' coarse-grained graph is equal to the MFPT between the vertices of the line connecting the subgraphs. { Cayley trees, the T-graph, $c$-ary trees {\it etc.} -- being trees -- all fall into this class.}
To the best of our knowledge, conservation of MFPTs under LE coarse-graining for the necklace type of graphs was previously unknown. 

We capitalise on the LE approach in two ways: 
(i) for the necklace class of graphs we are able to generalise the essential edge lemma (EEL) to non-reversible walkers 
and produce explicit and exact formulae for the MFPTs in cases where
the EEL is inapplicable.
(ii) The LE approach provides accurate and explicit (albeit approximate) formulae for MFPTs on graph structures that are not exact necklaces.

Explicit formulae in terms of network parameters for MFPTs are hard to come by: our LE approach offers a way to outmanoeuvre the infamous matrix inversion in eq.~\eqref{eq:mfpt_grounded_laplacian}, which in most cases could only be tackled numerically. Applications where explicit, although approximate, formulae are called for abound, as MFPTs are used as a low-order quantitative indicator in many different contexts. There is for instance an interest in MFPTs and FPTs to evaluate search strategies and transport for random walks, and models for diffusion on complex media (\cite{Benichou2014} and references therein). Further fields include the description of ill-mixed  gene regulatory network models \cite{Coulier2021} and kinetics of reactions in high-dimensional potentials \cite{Kells2020}. MFPTs have also been recently applied to assess the heterogeneity of complex social systems \cite{Bassolas2020}. Moreover, \cite{Kannan2020} show that the numerical error incurred using eq.~\eqref{eq:mfpt_grounded_laplacian} -- or other theoretically exact but numerically expensive methods -- may lead to large errors for MFPTs between different communities of vertices.

We have checked our exact analytical formulae against well-known results (or limits thereof) where available, or against results obtained by the established standard formula eq.~\eqref{eq:mfpt_grounded_laplacian}. All approximate results have been tested using numerical simulations, finding better agreement the closer the graphs
match the structure assumed in our method.

There are several interesting pathways for future work. 
First, in this work we have only considered graphs that 
could be coarse-grained into a one-dimensional lattice.
However, one may envisage to extend this framework to graphs that can be coarse-grained into loop-less graphs, i.e. unbalanced trees. 

Secondly, we have shown that the LE coarse-graining method provides reasonable approximations in test cases that deviate from the necklace structure, however 
it would be interesting to investigate the performance  
of the method more extensively, on different graph ensembles with community structure, e.g. the stochastic block model \cite{Holland1983}. 
More generally, establishing error bounds around our approximations will constitute an important next step. {Moreover, the method will require extension for structures that cannot naturally be clustered into 
loop-less graphs, 
such as small-world models where ``long-distance'' edges may introduce loops between clusters.}

{I}n this work we have focused on nodes along the 'backbone' of graphs with necklace structure. 
MFPTs between nodes residing 'far' from the backbone may be poorly described by the LE coarse-graining and other frameworks may be better suited for them. Recently, it has been shown that a coarse-graining method which was proposed by 
Hummer and Szabo \cite{Hummer2015} preserves the {\em average} MFPTs between clusters \cite{Kells2020}. This coarse-graining may provide more accurate estimates for nodes residing far from the backbone, however it leads to a more complex relation between the transition matrices of the original and clustered network.
It would be interesting to see whether analytical progress can be made for graph structures that allow one-dimensional coarse-grained representations, for such coarse-graining protocols. 

{Finally, we have focused entirely on \emph{mean} first-passage times. Higher moments and full distributions of FPTs are considerably less tractable than their mean, such that only specialised results limited to certain moments or as approximations in specific problems are available \cite{Bapat2011,Condamin2005}. However, numerical simulations presented in app.~\ref{app:higher_moments} suggest that our LE-coarse graining method might preserve higher moments of FPTs approximately if the graph is an exact necklace.}

\begin{acknowledgments}
PV and ET acknowledge support from UKRI Future Leaders Fellowship scheme [n. MR/S03174X/1]. Y-PF is supported by the EPSRC Centre for Doctoral Training in Cross-disciplinary Approaches to Non-Equilibrium Systems (CANES EP/L015854/1).
\end{acknowledgments}



\appendix

\section{Spanning trees, spanning forests and essential edges}\label{app:spanning_trees}

As explained in sec.~\ref{sec:main_derivation}, one can calculate the MFPTs on a graph by solving the combinatorial problem of finding all the spanning trees and forests of certain kinds (i.e. two-tree forests)
in the graph (see {eq.~}\eqref{eq:mfpt_tree_formula}). Below we provide the definitions of spanning trees and forests for the reader who is not familiar with these concepts. 

\begin{figure}[h]
    \centering
    \includegraphics{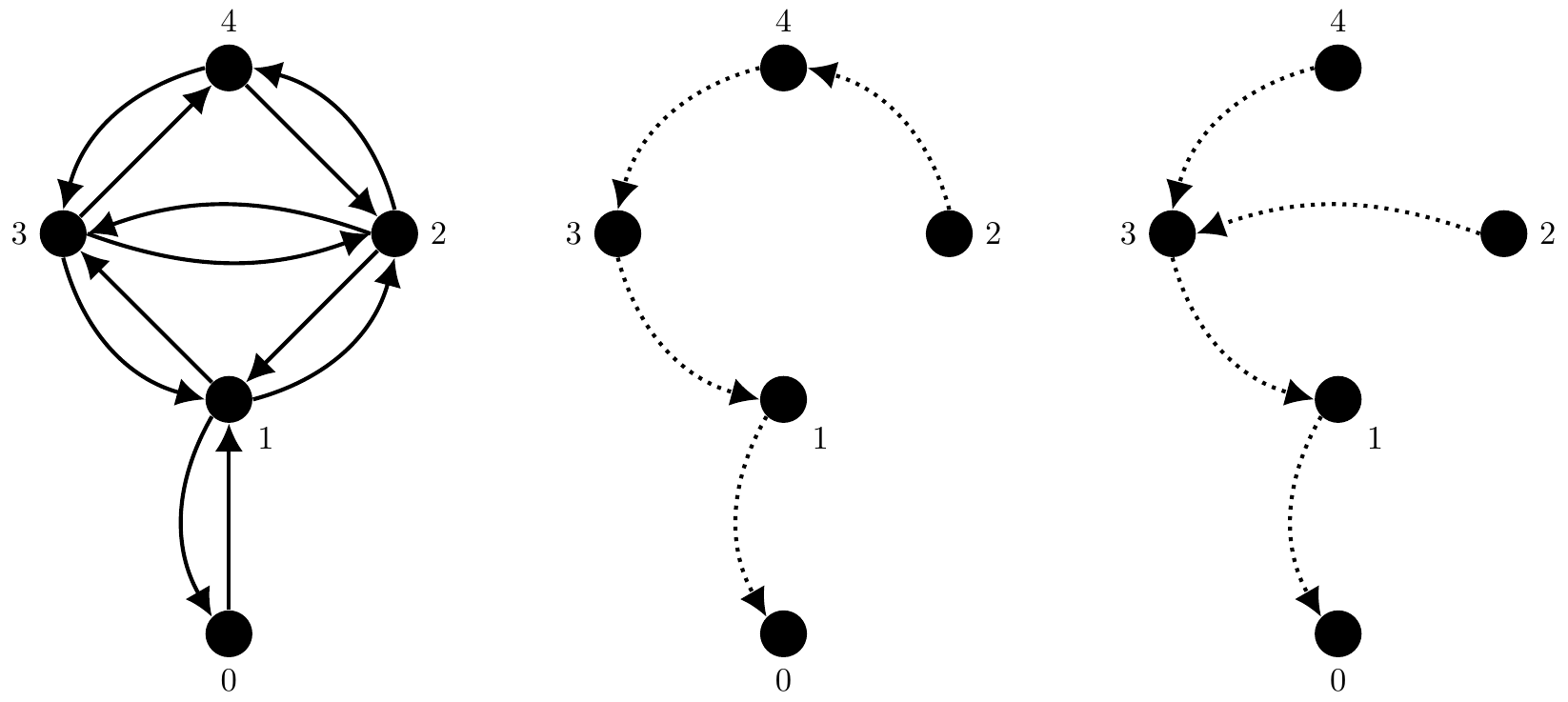}
    \caption{     
    Directed graph (left panel) and two
    examples of spanning trees with root $0$ (middle and right panels). 
    Note that the undirected edge ${(0,1)}$ is essential: Every spanning tree has to contain either $0\to 1$ or $1\to 0$, depending on the root.}
    \label{fig:spanning_trees}
 \end{figure}

\begin{figure}
    \centering
    \includegraphics{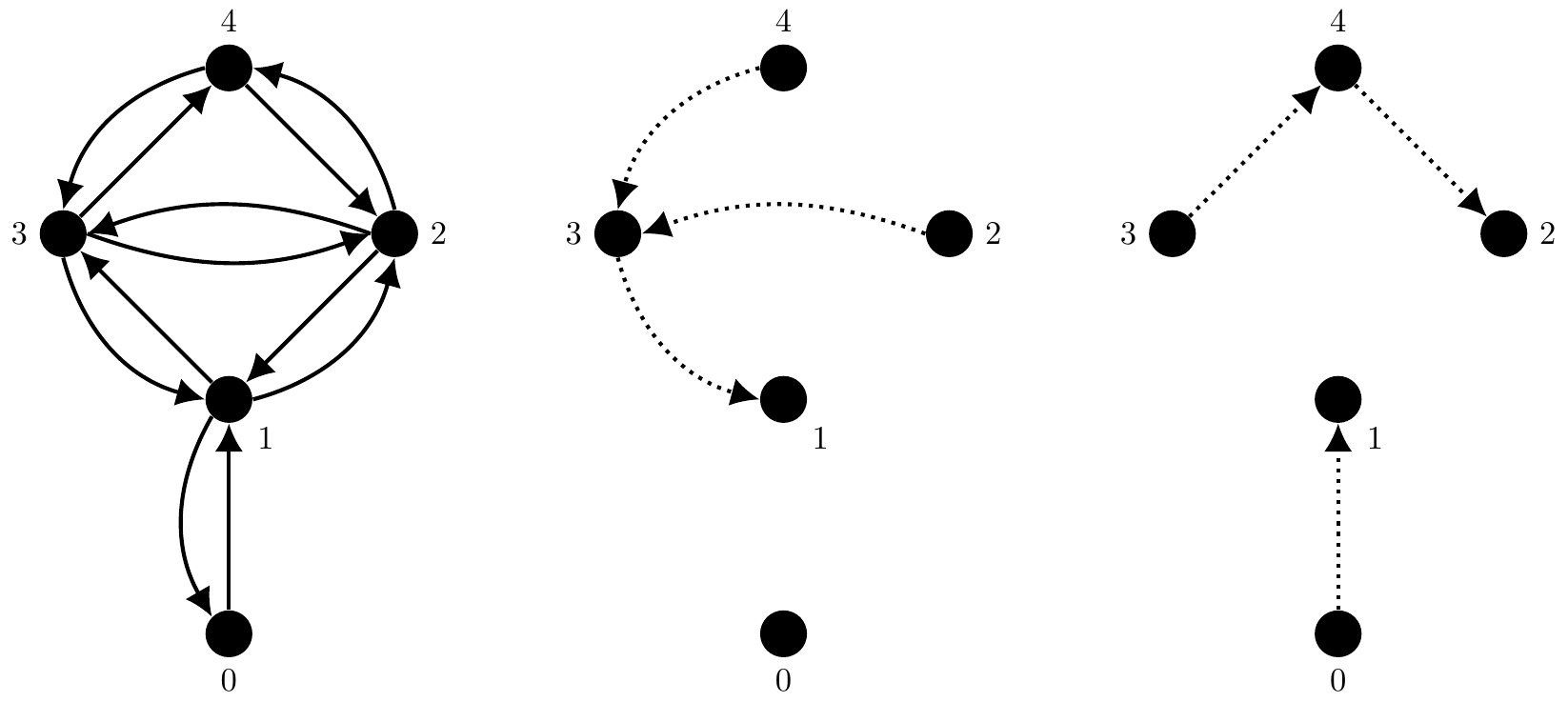}
    \caption{ 
    Directed graph (left panel) and two examples of spanning forests (middle and right panels).
    The middle forest has roots $0$ and $1$, the right forest has roots $1$ and $2$.}
    \label{fig:spanning_forests}
 \end{figure}

Given a directed graph, one defines a \emph{spanning forest} as a loopless directed subgraph that covers all vertices, while every vertex has at most one outgoing edge.
Those vertices without outgoing edges are the \emph{roots} of the forest. There is always at least one root, and if there are several, they define the different components, or trees, of the forest. In particular, a single-component spanning forest is called a \emph{spanning tree}.
For instance, fig.~\ref{fig:spanning_trees} shows two different directed spanning trees with root $0$. Similarly, fig.~\ref{fig:spanning_forests} shows two directed spanning forests with roots $\{0,1\}$ and $\{1,2\}$, respectively.

The problem of finding 
the spanning trees and two-tree forests of a graph is in general a formidable combinatorial task for large graphs. However, it may become feasible for special graph structures. 
For instance, if the graph itself has the structure of a tree, it will possess exactly one spanning tree for every root. 

{
\section{Higher Moments of FPTs}\label{app:higher_moments}
}

{
In this section we present some evidence for a generalisation of eq.~\eqref{eq:mfpt_LE_conservation_all} to higher moments, and in fact full distributions, of first-passage times (FTPs). To this end, let the FPT $t_{ij}$ be the first time step at which the walker is in state $j$, after having started from state $i$. We denote the $m$-th moment of $t_{ij}$ by
\begin{equation}\label{eq:FPT_moments}
    \lambda^m_{ij}:=\Mean(t^m_{ij})=\sum_{s=0}^{\infty} s^m \prob{t_{ij}=s}
\end{equation}
for $m\geq 0$. Evidently, $\lambda^0_{ij}=1$ due to normalisation, and $\lambda^1_{ij}=m_{ij}$ by definition of the MFPT from $i$ to $j$.
}

{
For the first-passage process to $j$, we may without loss of generality assume that $j$ is an absorbing state. In that case, the $j$-th row of the transition matrix $\bq$ has a unit entry in the $j$-th column and $0$ everywhere else. The probability that $t_{ij}\leq s$ is then given by the probability that the walker is in state $j$ at time $s$ (as it can have entered $j$ either before time $s$ and never left, or entered at time $s$). In terms of $\bq$ this reads
\begin{equation}
    \prob{ t_{ij}\leq s }=\left(\bq^{s}\right)_{ij}\ ,
\end{equation}
or for the probability mass function (PMF)
\begin{equation}\label{eq:FPT_pmf}
    \prob{ t_{ij} = s} =\left(\bq^{s}\right)_{ij}-\left(\bq^{s-1}\right)_{ij}\ .
\end{equation}
Analogously, we denote FPTs and their moments on the coarse-grained graph by $T_{IJ}$ and $\Lambda_{IJ}^m$, respectively.
}

{
Led by eq.~\eqref{eq:mfpt_LE_conservation_all}, we now test if on a necklace the original and coarse-grained walker have the same FPT distributions or moments, i.e. if $\prob{t_{v_0 v_H}=s}=\prob{T_{0H}=s}$ or $\lambda^m_{v_{I-1} v_I}=\Lambda^m_{I-1, I}$ for any $m>1$. We do this by way of example, considering a necklace of five cliques with five vertices, each, following the examples in sec.~\ref{sec:cliques}. Additionally, every edge is weighted by a number drawn independently and uniformly from the unit interval, and both edge directions are taken to be independent as well (i.e. the edges $(i,j)$ and $(j,i)$ are both present and weighted independently for each edge $(i,j)$ present in the necklace as per sec.~\ref{sec:cliques}). 
}

{
For a single realisation of edge weights, the PMFs obtained using eq.~\eqref{eq:FPT_pmf} are shown in fig.~\ref{fig:fpt_pmf} along with their Kullback-Leibner divergence. To this end, both PMFs were truncated such that
\begin{equation}
    \prob{t_{v_0 v_H} \leq t_\textup{trunc}} \approx \prob{T_{0 H} \leq t_\textup{trunc}} \approx 0.9999  
\end{equation}
to avoid numerical problems with the Kullback-Leibler divergence when the PMFs range close to zero.
The two PMFs show an excellent agreement, with only a slight relative shift of $T_{0 H}$ to higher values. 
\begin{figure}[h]
    \centering
    \includegraphics{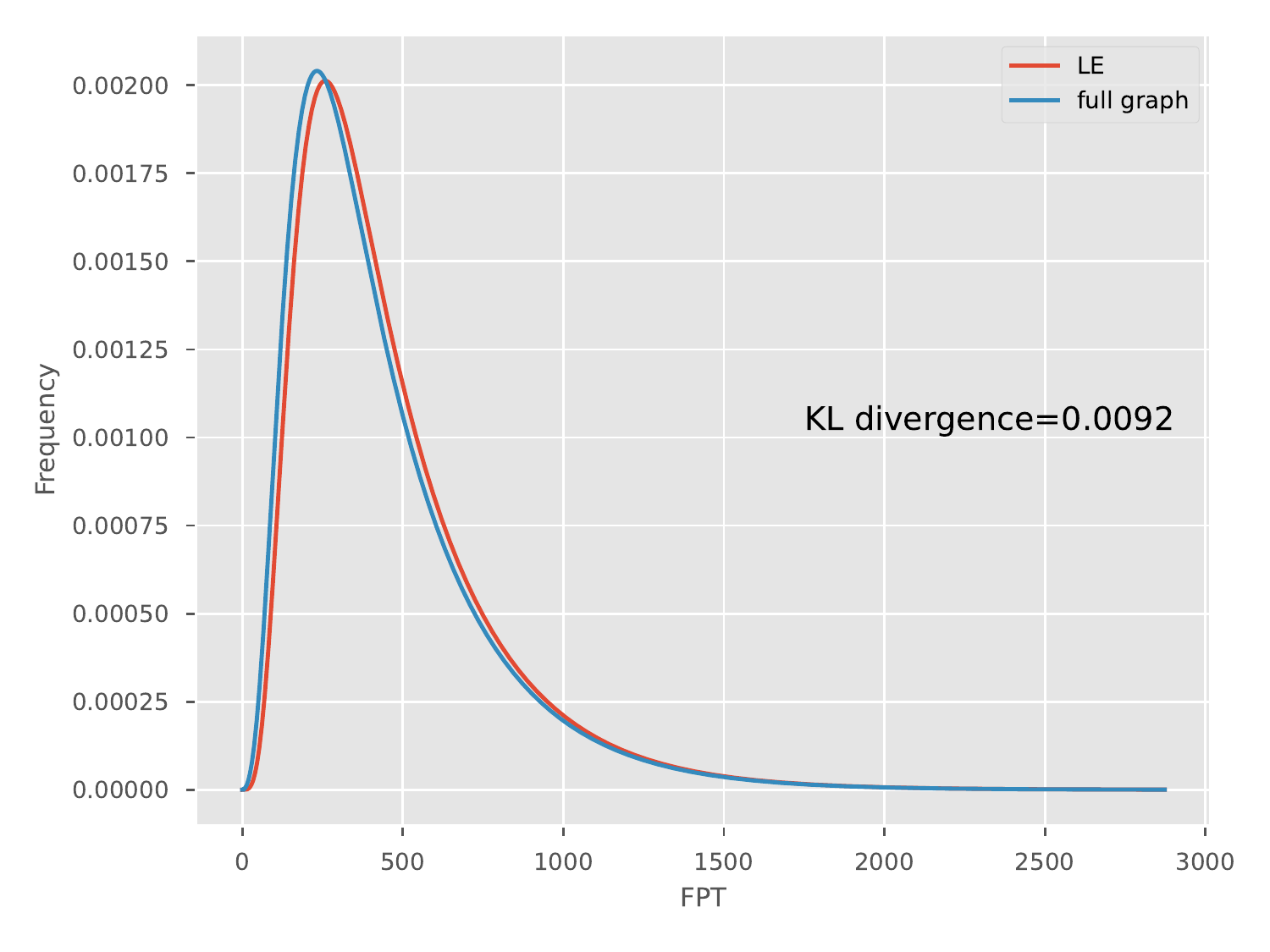}
    \caption{First-passage time PMFs for coarse-grained and full walker, obtained using eq.~\eqref{eq:FPT_pmf}. The shown part of the PMFs account for a fraction of $0.9999$ of the total mass, each.}
    \label{fig:fpt_pmf}
\end{figure}
}

{
Similarly, in fig.~\ref{fig:fpt_moments}, the root-moments $\sqrt[m]{\Lambda^m_{0 H}}$ of the coarse-grained walker are plotted against the root-moments $\sqrt[m]{\lambda^m_{v_0 v_H}}$ of the original walker for $m=2,3,4,5,10,15$ and over $200$ realisations of edge weights. The moments were computed by truncating the PMFs, eq.~\eqref{eq:FPT_pmf}, at $t=5,000$ and applying eq.~\eqref{eq:FPT_moments} (again truncated at $t=5,000$).
}

{
Fig.~\ref{fig:fpt_moments} shows an excellent -- though not exact -- agreement between the moments $\Lambda^m_{0 H}$ and $\lambda^m_{v_0 v_H}$. There is a substantial disagreement between the two only for comparatively low values of $\lambda^m_{v_0 v_H}$, where $\Lambda^m_{0 H}$ is higher in a few cases.
\begin{figure}[h]
    \centering
    \includegraphics{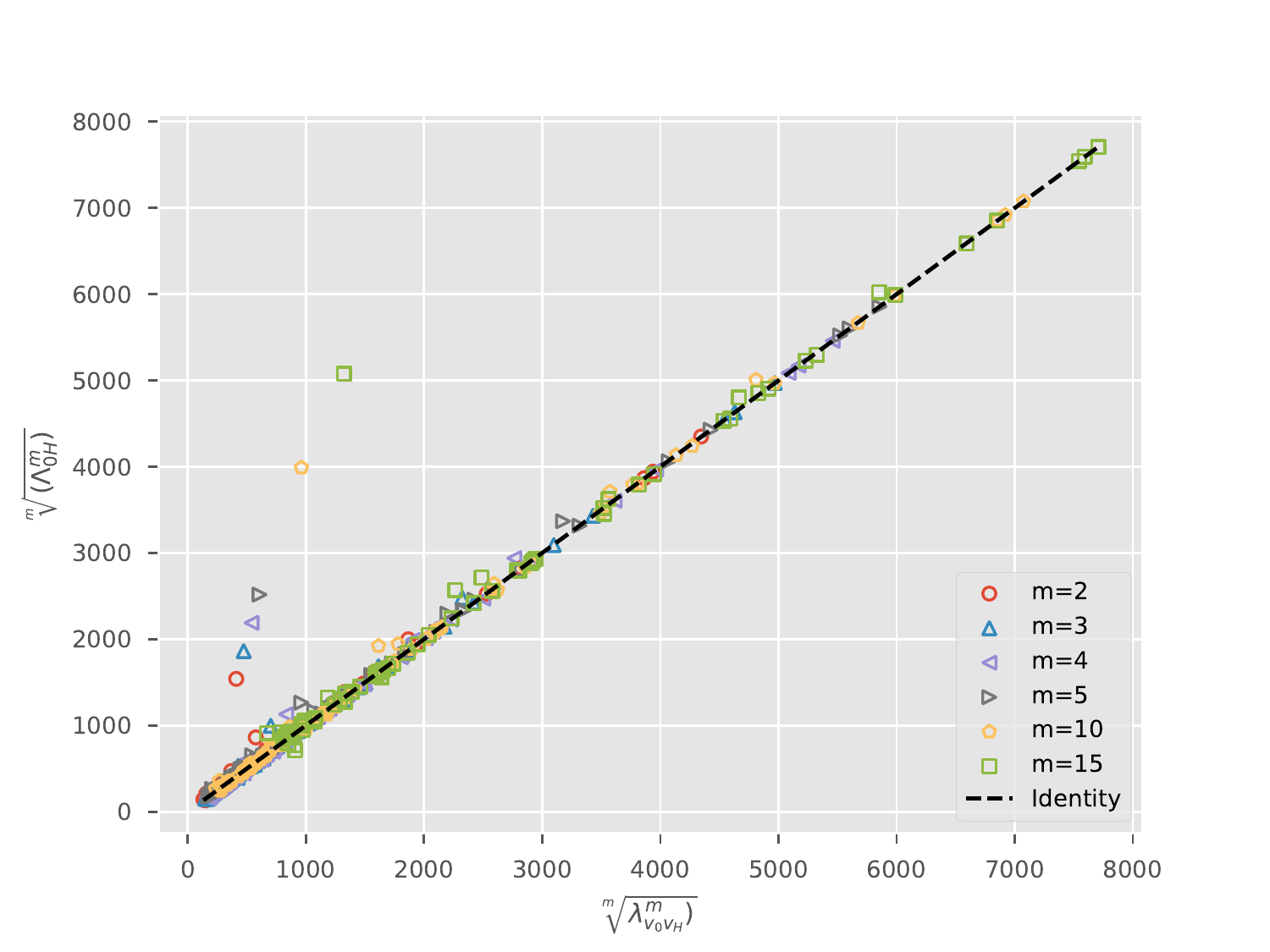}
    \caption{FPT root-moments of the coarse-grained and the full walker for $200$ realisations of edge weights.}
    \label{fig:fpt_moments}
\end{figure}
}

\end{document}